\documentclass[
  aps,
  pra,
  reprint,
  superscriptaddress,
  amsmath,
  amssymb,
  longbibliography
]{revtex4-2}

\usepackage{hyperref} 
\usepackage{cleveref}
\usepackage{crossreftools}
\crefname{equation}{eq.}{eqs.}
\Crefname{equation}{Eq.}{Eqs.}
\crefname{figure}{fig.}{figs.}
\Crefname{figure}{Fig.}{Figs.}

\usepackage{tikz}
\usetikzlibrary{math}
\usepackage{url}
\usepackage{seqsplit}
\usepackage{xstring}
\usepackage[utf8]{inputenc}
\usepackage{graphicx}
\usepackage{physics}
\usepackage{braket}
\usepackage{pdfpages}

\makeatletter
\AtBeginDocument{\let\LS@rot\@undefined}
\makeatother

\newcommand{\Z}[1]{\langle Z^{#1}_{\rm tot}\rangle}
\newcommand{\X}[1]{\langle X^{#1}_{\rm tot}\rangle}

\newcommand{\FigMPSDrawing}{
\begin{figure}[t!]
  \centering
  \includegraphics[width=\columnwidth]{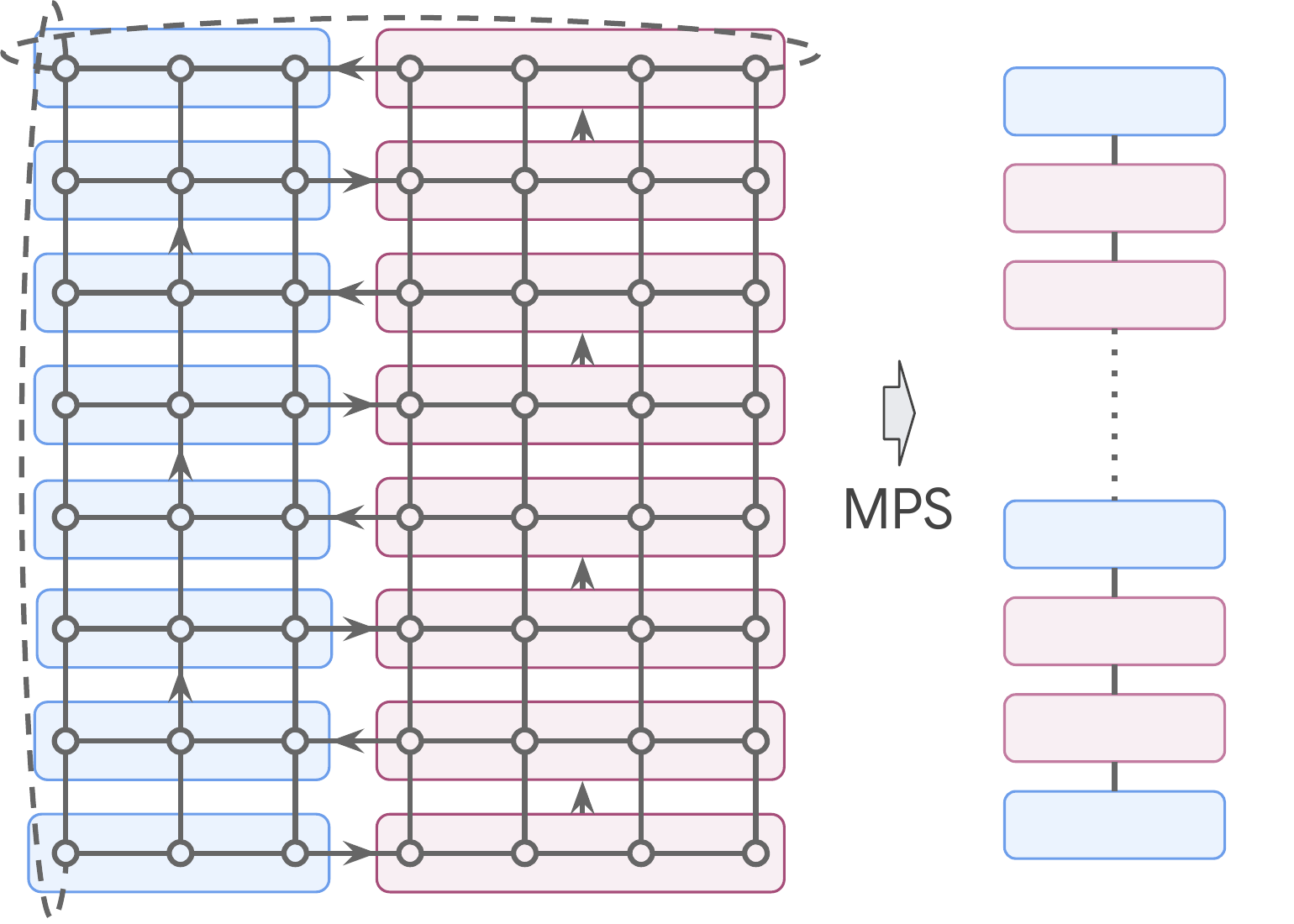}
  \caption{
      \textbf{(Left) Example of qubit grouping for a $\boldsymbol{7\times8}$
      grid. (Right) Final structure of the MPS.} Qubits in each row are split
      into two groups, and qubits in the same group are fused together. The
      dashed lines correspond to the boundary conditions, while the arrows
      represent the ordering of the blocks in the MPS.
  }
  \label{fig:mps_drawing}
\end{figure}
}

\newcommand{\FigFiveBySixZ}{
\begin{figure}[t!]
    \centering
    \includegraphics[width=\columnwidth]{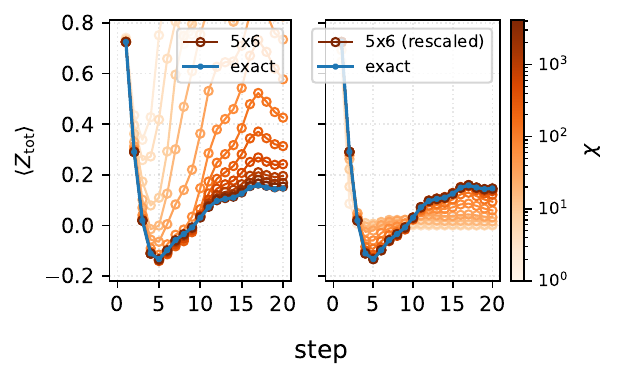}\\
    \includegraphics[width=0.9\columnwidth]{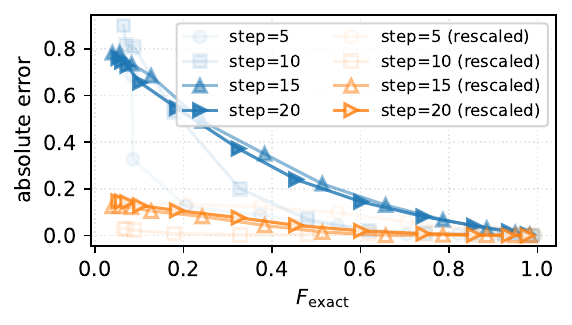}
    \caption{
        \textbf{MPS computation of the magnetization, at fixed
        $\boldsymbol{(\Delta\theta,\, J,\, h,\, dt) = (2\pi/9, \,-1, \,2,
        \,0.25)}$.} The top plots show the convergence of $\Z{}_{\rm MPS}$
        without rescaling (left) and with rescaling using the fidelity $F_{\rm
        MPS}$ (right), see \Cref{eq:Z_tot_via_F}. The bottom plot shows
        the absolute error at a fixed step as a function of the true fidelity
        $F_{\rm exact} = |\langle \psi_{\rm exact} | \psi_{\rm MPS} \rangle|^2$,
        for both $\Z{}_{\rm MPS}$ (filled blue markers) and $\Z{}_{\gamma_{\rm
        fit}}$ (open orange markers).
    }
    \label{fig:Z_tot_5x6}
\end{figure}
}

\newcommand{\FigConvergence}{
\begin{figure}[t!]
    \centering
    \includegraphics[width=\columnwidth]{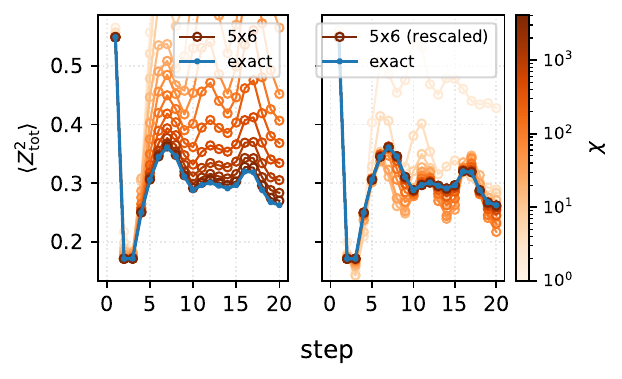}\\
    \includegraphics[width=0.9\columnwidth]{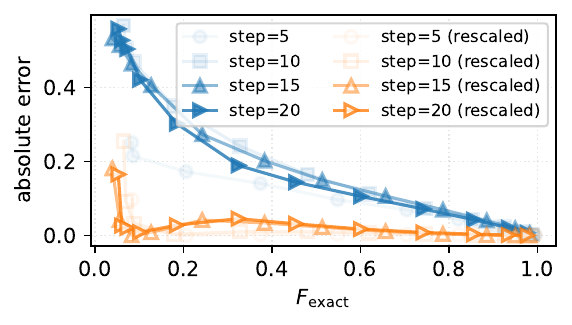}
    \caption{
        \textbf{Comparison of the Ising order parameter with and without
        rescaling, at fixed $\boldsymbol{(\Delta\theta,\, J,\, h,\, dt) =
        (2\pi/9, \,-1, \,2, \,0.25)}$.} The top-left and top-right plots show
        the convergence of $\Z{2}_{\rm MPS}$ and $\Z{2}_{\gamma_{\rm fit}}$,
        respectively, with increasing bond dimension. The bottom plot shows
        the absolute error at a fixed step as a function of the true fidelity
        $F_{\rm exact} = |\langle \psi_{\rm exact} | \psi_{\rm MPS} \rangle|^2$,
        for both $\Z{2}_{\rm MPS}$ (filled blue markers) and $\Z{2}_{\gamma_{\rm
        fit}}$ (open orange markers).
    }
    \label{fig:convergence}
\end{figure}
}

\newcommand{\FigGammaScaling}{
\begin{figure}[t!]
    \centering
    \includegraphics[width=\columnwidth]{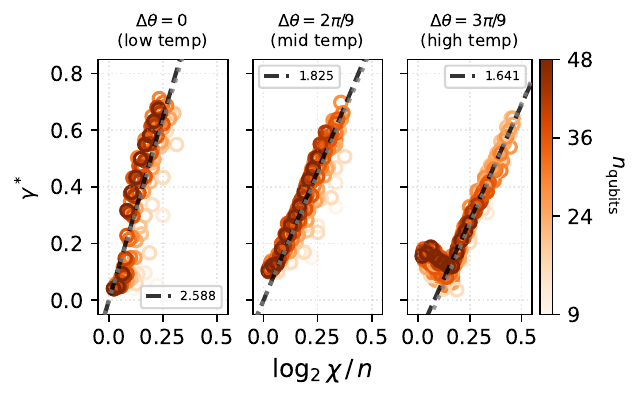}
    \caption{
        \textbf{Fit for the $\boldsymbol{\gamma^*}$ exponent for three different
        temperature.} Correlation between the optimal exponent $\gamma^*$ and
        the rescaled bond dimension $\log_2\chi / n$, with varying system size
        and shape. The gray-dotted lines correspond to the scaling of $\gamma^*$
        if a one-to-one mapping were used instead of grouping sites together.
        The fit is computed using systems with between $20$ and $36$ qubits.
        The fitting parameters are reported in \Cref{tab:gamma_scaling}.
    }
    \label{fig:gamma_scaling}
\end{figure}
}

\newcommand{\FigRescalingError}{
\begin{figure}[t!]
    \centering
    \includegraphics[width=\columnwidth]{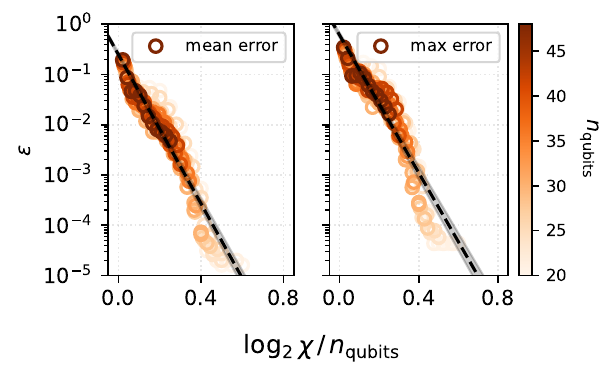}
    \caption{
        \textbf{Scaling of the absolute error between the rescaled
        $\boldsymbol{\Z{2}_{\gamma_{\rm fit}}}$ and the exact
        $\boldsymbol{\Z{2}_{\rm exact}}$, at fixed
        $\boldsymbol{(\Delta\theta,\, J,\, h,\, dt) = (2\pi/9, \,-1, \,2,
        \,0.25)}$.} The plot shows the mean absolute error (Left) and the
        maximum absolute error (Right) between $\Z{2}_{\gamma_{\rm fit}}$ and
        $\Z{2}_{\rm exact}$ over $s = 20$ steps, for different grid sizes and
        bond dimensions $\chi$. The regression $\epsilon = K \, \chi^{\alpha/n}$
        is computed by considering only instances with at least $20$ qubits (see
        \Cref{tab:scaling_error_params} for the fitting parameters). The shaded
        area corresponds to the linear regression error. See also
        \Cref{fig:TFIM_rescaling_error} in the Supplemental Material.
    }
    \label{fig:rescaling_error}
\end{figure}
}

\newcommand{\FigZvsZZ}{
\begin{figure*}[t!]
    \centering
    \includegraphics[width=0.331\textwidth]{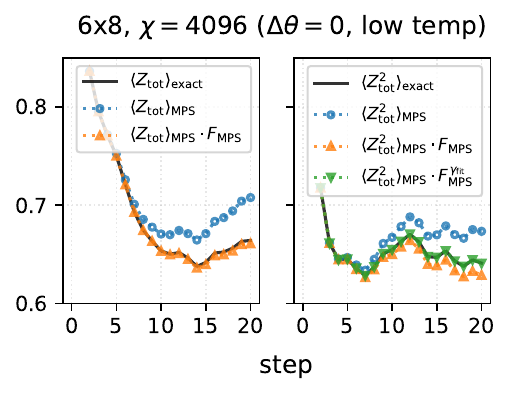}
    \hspace{-15px}
    \includegraphics[width=0.35\textwidth]{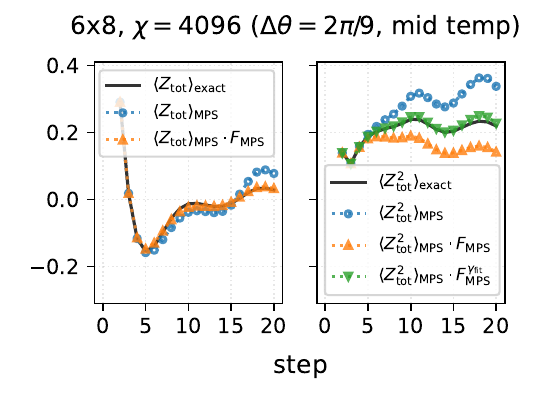}
    \hspace{-18px}
    \includegraphics[width=0.35\textwidth]{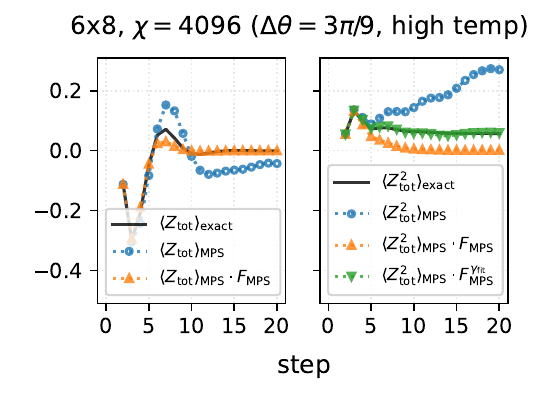}
    \caption{
        \textbf{Comparison of the single-body observable $\boldsymbol{\Z{}}$
        with $\boldsymbol{\Z{2}}$, at different temperatures.} Here, the
        $\gamma_{\rm fit}$ values are $0.647$, $0.456$, and $0.280$ for low-,
        mid-, and high-temperature, respectively.
    }
    \label{fig:z1_vs_z2}
\end{figure*}
}

\newcommand{\FigSixByEightLowMidHighTemp}{
\begin{figure*}[t!]
    \centering
    \includegraphics[width=\textwidth]{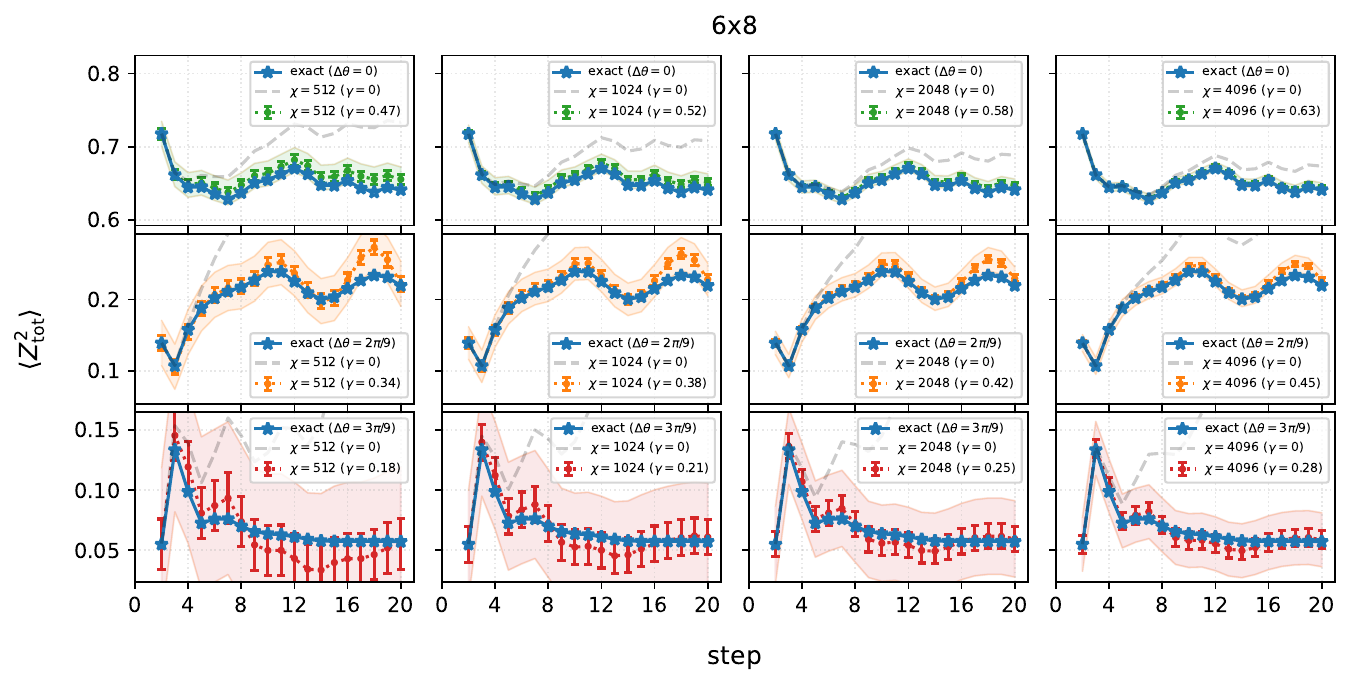}
    \caption{
        \textbf{Numerical simulations of a $\boldsymbol{6\times8}$ TFIM for
        three different initial temperatures, by varying the maximum bond
        dimension $\boldsymbol{\chi}$.} For each temperature, we identified the
        optimal $\gamma_{\rm fit}$ and the corresponding extrapolated error,
        following the procedures detailed for the mid-temperature case, as shown
        in \Cref{fig:gamma_scaling} and \Cref{fig:rescaling_error}. The
        blue-starred lines represent the exact results. The orange-circled lines
        correspond to our numerical results for $\Z{2}_{\gamma_{\rm fit}}$, with
        the error bars (shaded areas) corresponding to the estimates from the
        mean (maximum) absolute error from \Cref{fig:rescaling_error}. For
        comparison, the gray-dashed lines correspond to the value of $\Z{2}_{\rm
        MPS}$ without any rescaling. For all temperatures, $\Z{2}_{\gamma_{\rm
        fit}}$ is within the maximum error bound of $\Z{2}_{\rm exact}$.
    }
    \label{fig:6x8_z2_low_mid_high_temp}
\end{figure*}
}

\newcommand{\FigSevenByEightZZ}{
\begin{figure*}[t!]
    \centering
    \includegraphics[width=0.9\textwidth]{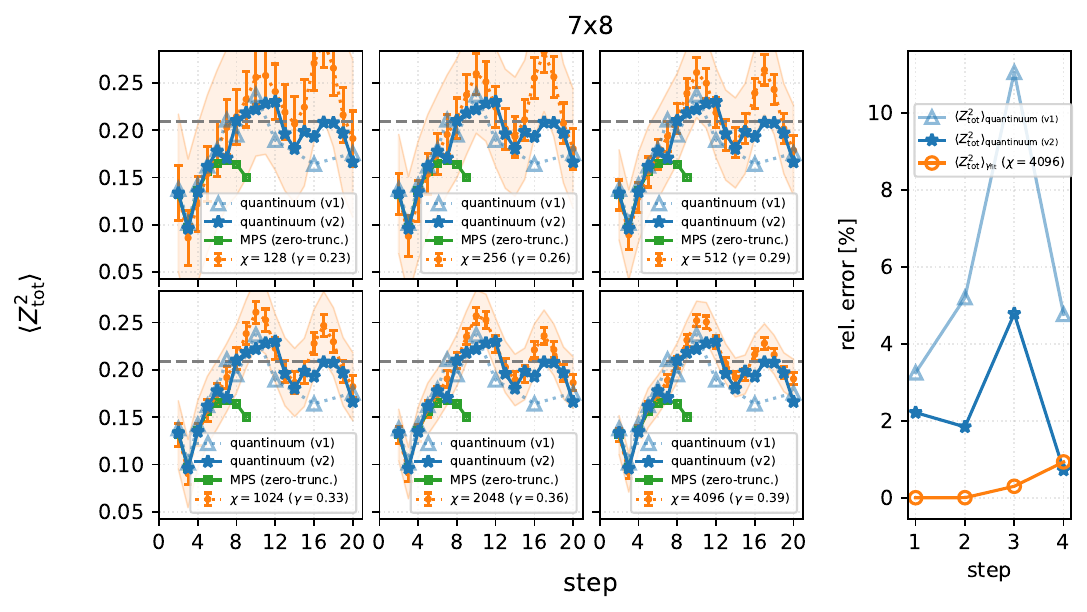}
    \caption{
        \textbf{(Left) Numerical simulations of a $\boldsymbol{7\times8}$ TFIM,
        by varying the maximum bond dimension $\boldsymbol{\chi}$. (Right)
        Comparison against exact results.} The blue-starred and blue-triangled
        lines represent the results from the Quantinuum H2 device before
        \cite{Haghshenas2025Digital_v1} and after
        \cite{Haghshenas2025Digital} the implementation of their improved error
        correction protocol.
        We also report the zero-truncation MPS results from
        \cite{Haghshenas2025Digital}. The orange-circled lines correspond to our
        numerical results for $\Z{2}_{\gamma_{\rm fit}}$, with the error bars
        (shaded areas) corresponding to the estimates of the mean (maximum)
        absolute error from \Cref{fig:rescaling_error}. The dashed horizontal
        line corresponds to the expected thermal value of $\Z{2}$ using an MPS
        purification ansatz (from \cite{Haghshenas2025Digital}).
    }
    \label{fig:7x8_z2}
\end{figure*}
}

\newcommand{\FigLargeGrids}{
\begin{figure*}[t!]
    \centering
    \includegraphics[width=\textwidth]{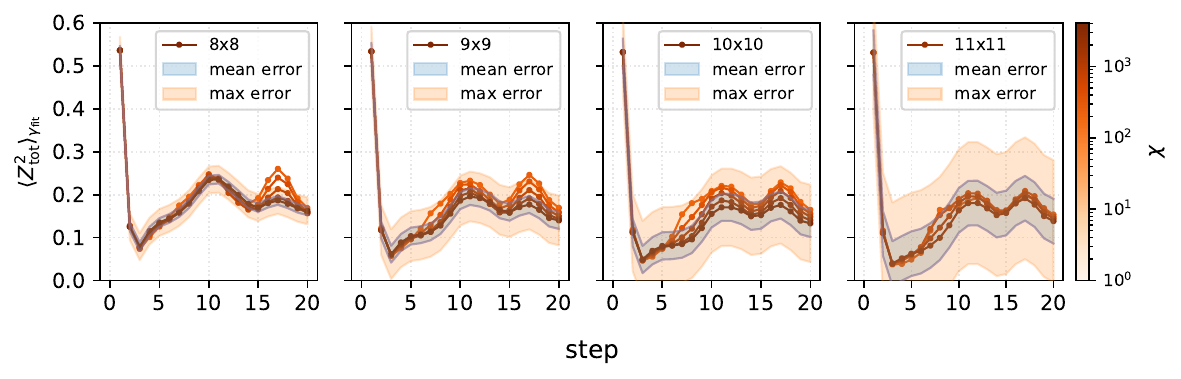}
    \caption{
        \textbf{Numerical results for $\boldsymbol{\Z{2}_{\gamma_{\rm fit}}}$
        for larger grids, at fixed $\boldsymbol{\Delta\theta = 2\pi/9}$
        (mid-temperature).} The plots show the rescaled $\Z{2}_{\gamma_{\rm
        fit}}$ for different grid sizes, with varying maximum bond dimension
        (only bond dimensions with $\chi \geq 256$ are visualized). The largest
        bond dimension for all grids except $11\times11$ is $\chi = 4096$. For
        the $11\times11$ grid, the maximum bond dimension is $\chi = 2048$. The
        shaded areas correspond to the extrapolated mean absolute error and the
        extrapolated maximum absolute error for the largest available bond
        dimension, obtained from \Cref{fig:rescaling_error}.
    }\label{fig:large_grids}
\end{figure*}
}

\newcommand{\FigThermal}{
\begin{figure}[t!]
    \centering
    \includegraphics[width=\columnwidth]{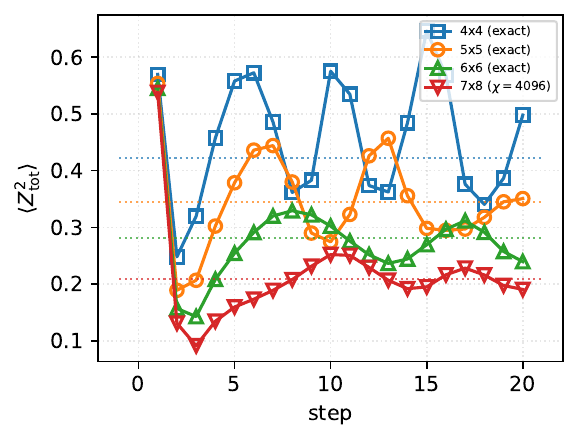}
    \caption{
        \textbf{Expected thermal value of $\boldsymbol{\Z{2}}$ for the Floquet
        Hamiltonian computed in the canonical ensemble using an MPS purification
        ansatz (from  \cite{Haghshenas2025Digital}), compared to either
        $\boldsymbol{\Z{2}_{\rm exact}}$ or $\boldsymbol{\Z{2}_{\gamma_{\rm
        fit}}}$.} Our numerical $\Z{2}_{\gamma_{\rm fit}}$ for the $7\times8$
        system oscillates around the expected thermal value, corroborating its
        validity. Results for the expected thermal values are from
        \cite{Haghshenas2025Digital}.
    }
    \label{fig:thermal}
\end{figure}
}

\newcommand{\FigOrthogonalConvergence}{
  \begin{figure}[t!]
    \centering
    \includegraphics[width=\columnwidth]{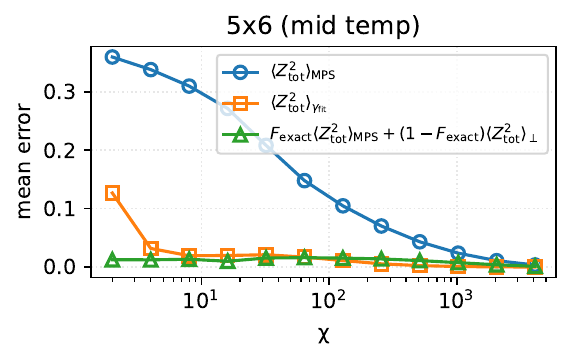}
    \caption{
      \textbf{Mean absolute error of the Ising order parameter as a function of
      the bond dimension $\boldsymbol{\chi}$.} The plot shows the absolute
      error, averaged over 20 Trotter steps, for the Ising order parameter:
      without rescaling ($\Z{2}_{\rm MPS}$), with rescaling via the fidelity
      ($\Z{2}_{\gamma_{\rm fit}}$), and with the inclusion of the discarded
      component after truncation ($F_{\rm exact} \Z{2}_{\rm MPS} + (1 - F_{\rm
      exact}) \Z{2}_\perp$). Here, $F_{\rm exact} = |\langle \psi_{\rm exact} |
      \psi_{\rm MPS} \rangle|^2$. To evaluate $|\psi_\perp\rangle$, we
      concurrently evolved the MPS and exact simulations, performing a
      Gram-Schmidt orthogonalization at the end of each Trotter step to extract
      $|\psi_\perp\rangle$ from $|\psi_{\rm exact}\rangle$.
    }
    \label{fig:orthogonal}
  \end{figure}
}

\newcommand{\FigXYComparison}{
  \begin{figure*}[htb]
    \centering
    \includegraphics[width=0.31\textwidth]{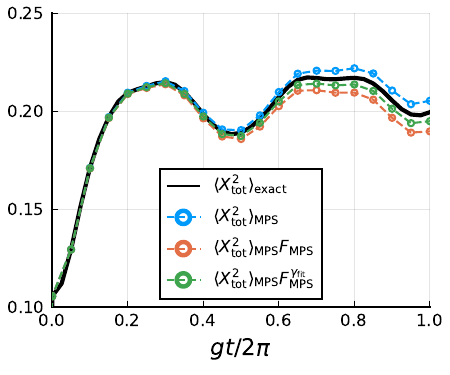}
    \includegraphics[width=0.31\textwidth]{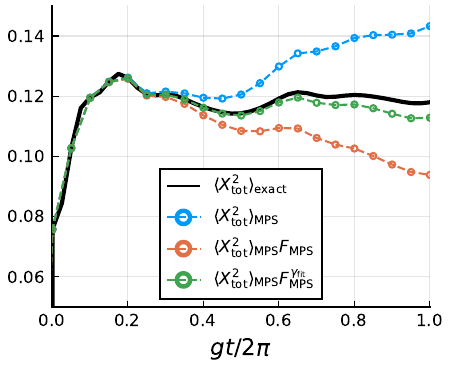}
    \includegraphics[width=0.31\textwidth]{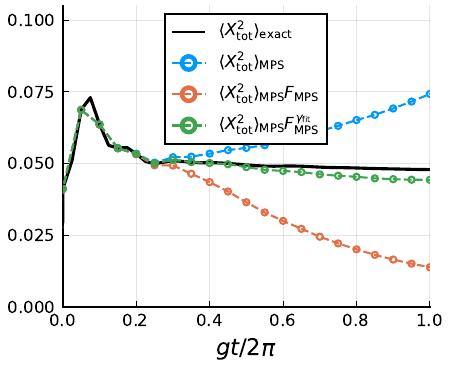}
    \caption{
      \textbf{Comparison of scaled and unscaled $\boldsymbol{\X{2}}$ to exact
        data, for different temperatures.} For all temperatures, curves scaled
      by a fractional power of fidelity agree much better with the exact data
      than either unscaled curves or curves scaled by fidelity. Here, the
      system is on a $6 \times 6$ grid and the MPS data is computed with
      $\chi=4096$. The states from left to right use the low-, intermediate-,
      and high-energy state preparations, respectively, as described in
      \Cref{sec:xy_stateprep}, and $\gamma$ is extracted from the fits reported in
      \Cref{fig:xy_gamma_scaling} of the Supplemental Material.
    }
    \label{fig:xy_6x6_4096}
  \end{figure*}
}

\newcommand{\FigXYLargeGrids}{
  \begin{figure*}[t!]
    \centering
    \includegraphics[width=\textwidth]{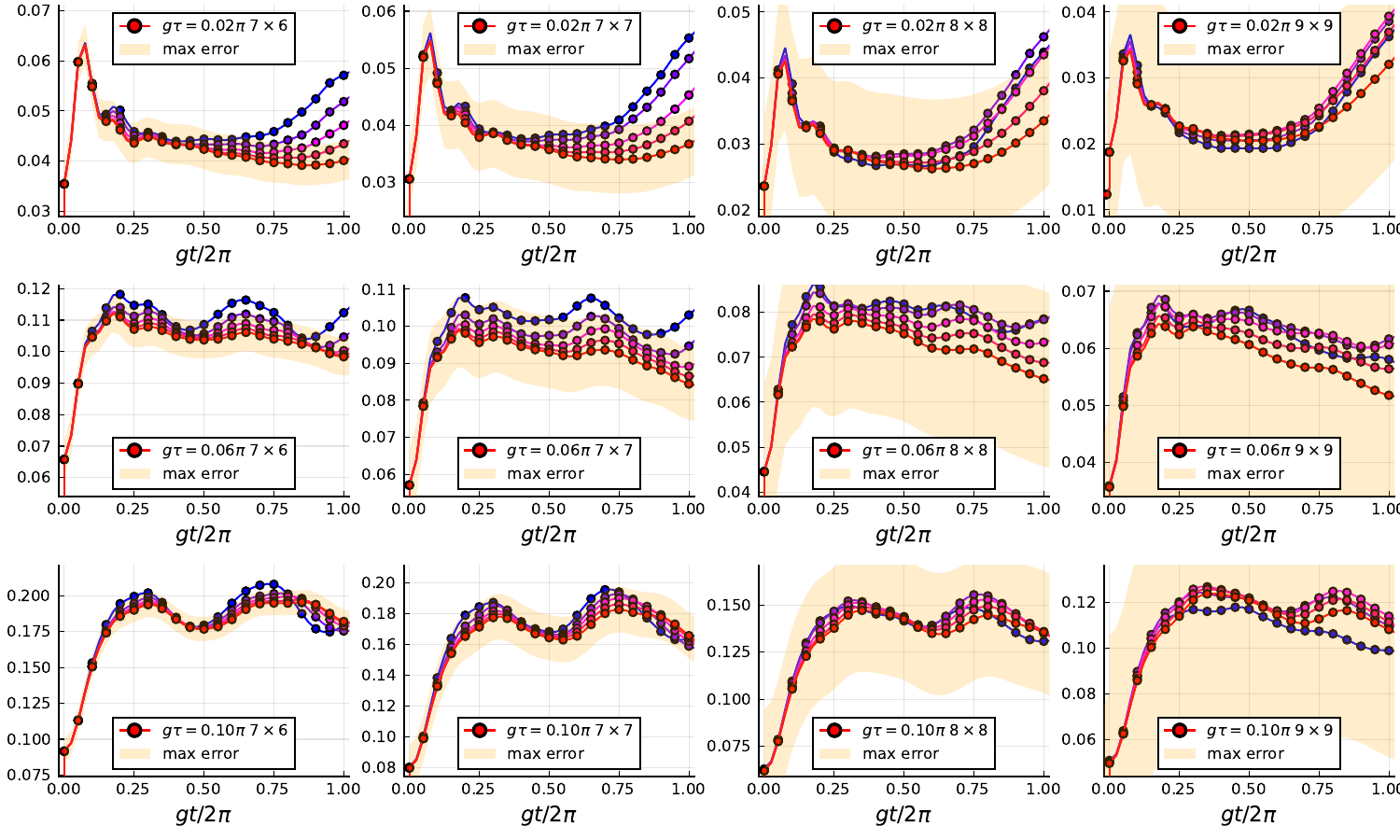}
    \caption{
      \textbf{Numerical results for the XY order parameter dynamics on larger
        grids.} The plots show the rescaled $\X{2}_{\gamma_{\rm fit}}$ for
      different grid sizes, with the maximum bond dimension ranging from $128$
      (blue) to $2048$ (red). The shaded areas correspond to the extrapolated
      maximum absolute error for a bond dimension of $2048$, as determined from
      the fits reported in \Cref{fig:xy_rescaling_error} of the Supplemental Material.
    }
    \label{fig:xy_large_grids}
  \end{figure*}
}

\newcommand{\FigRescalingErrorTwoD}{
  \begin{figure*}[h!]
    \centering
    \includegraphics[width=0.49\textwidth]{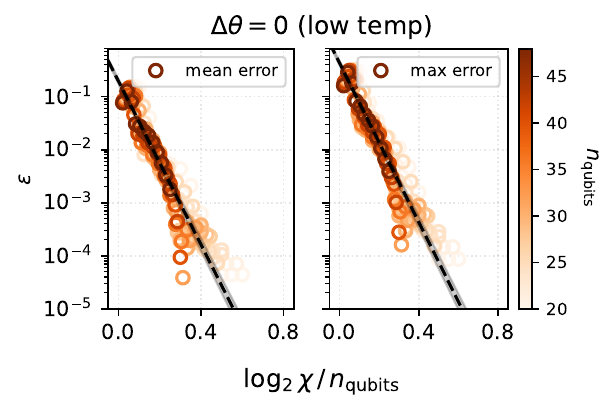}
    \includegraphics[width=0.49\textwidth]{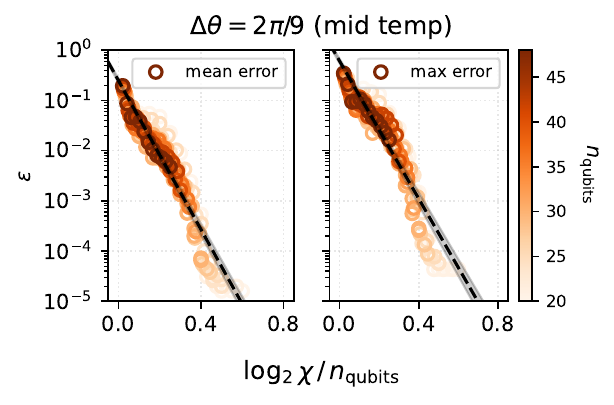}
    \includegraphics[width=0.49\textwidth]{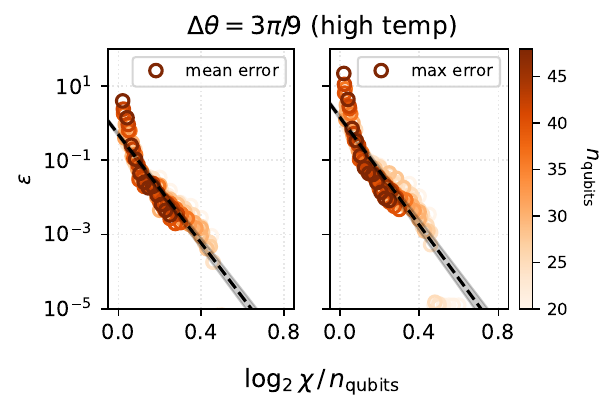}
    \caption{
      \textbf{Scaling of the absolute error between the rescaled
      $\boldsymbol{\Z{2}_{\gamma_{\rm fit}}}$ and the exact
      $\boldsymbol{\Z{2}_{\rm exact}}$ for the 2D TFIM, at fixed
      $\boldsymbol{(J,\, h,\, dt) = (-1, \,2, \,0.25)}$, across three
      different initial temperatures.} The plots show the mean absolute error
      (left) and the
      maximum absolute error (right) between $\Z{2}_{\gamma_{\rm fit}}$ and
      $\Z{2}_{\rm exact}$ over $s = 20$ steps for different grid sizes and bond
      dimensions $\chi$.  The regression $\epsilon = K \, \chi^{\alpha/n}$ is
      computed considering only instances with at least $20$ qubits (see
      \Cref{tab:scaling_error_params} of the main text for the fitting
      parameters). The shaded area corresponds to the linear regression error.
    }
    \label{fig:TFIM_rescaling_error}
  \end{figure*}
}

\newcommand{\FigSnake}{
  \begin{figure*}[h!]
    \centering
    \includegraphics[width=0.55\textwidth]{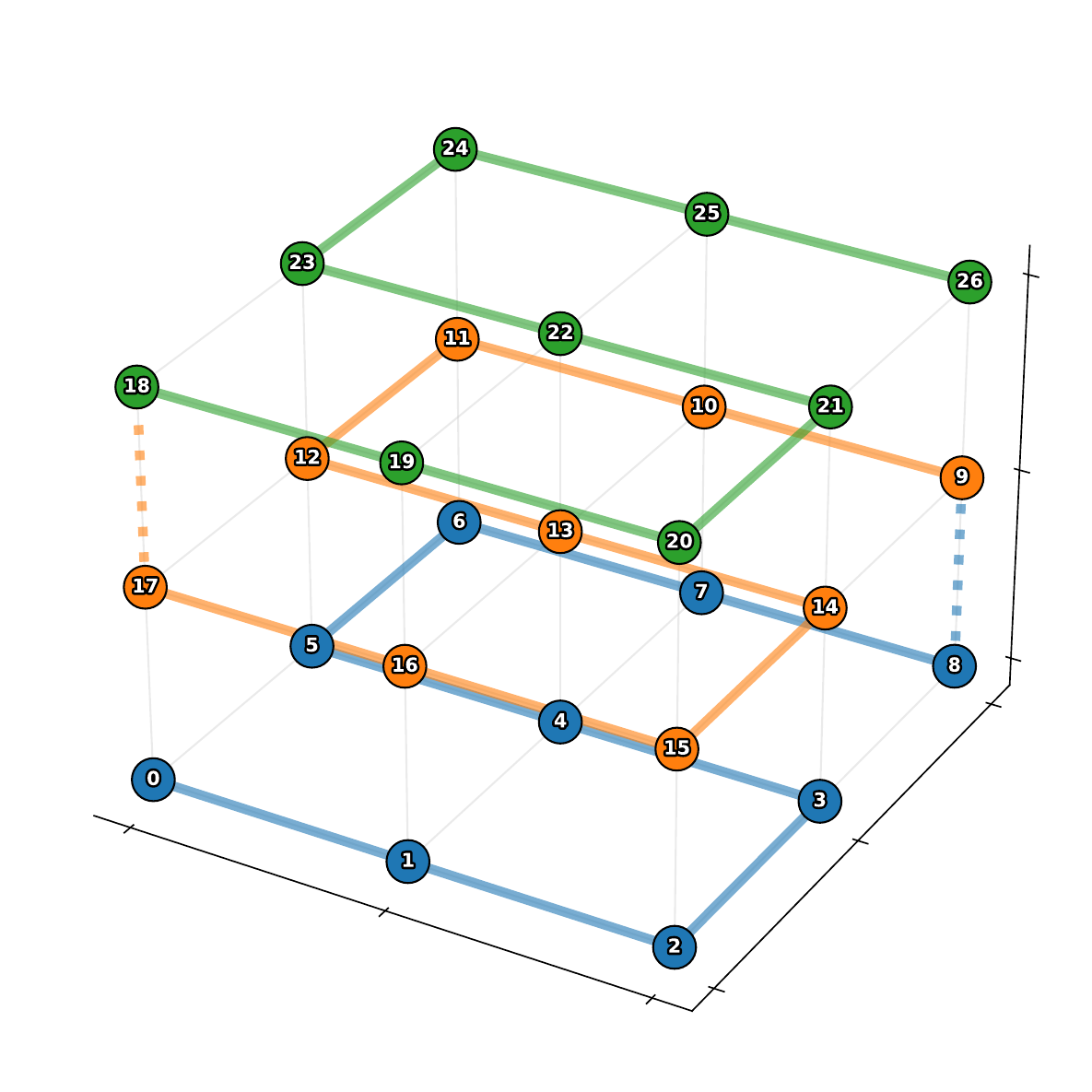}
    \caption{
      \textbf{Snake MPS ordering for a $\boldsymbol{3\times3\times3}$ grid.} The
      plot depicts the snake MPS ordering. The snake is colored differently for
      each level to improve readability, and the dashed lines represent the
      jumps in the snake order between levels.
    }
    \label{fig:3D_snake}
  \end{figure*}
}

\newcommand{\FigTFIMThreeDGammaScaling}{
  \begin{figure*}[h!]
    \centering
    \includegraphics[width=0.65\textwidth]{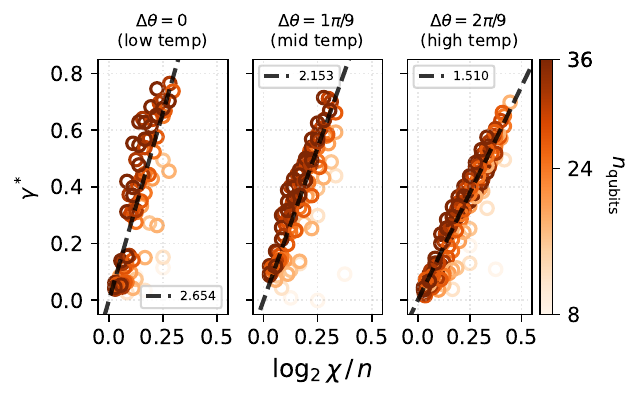}
    \caption{
      \textbf{Fit for the $\boldsymbol{\gamma^*}$ exponent for the 3D TFIM,
      across three different initial temperatures.} Correlation between the
      optimal exponent $\gamma^*$ and the rescaled bond dimension $\log_2\chi /
      n$ for various system sizes and shapes. The fit is computed using systems
      containing between $20$ and $32$ qubits. The fitting parameters are
      reported in \Cref{tab:3d_gamma_scaling}.
    }
    \label{fig:TFIM_3D_gamma_scaling}
  \end{figure*}
}

\newcommand{\FigTFIMThreeDRescalingError}{
  \begin{figure*}[h!]
    \centering
    \includegraphics[width=0.48\textwidth]{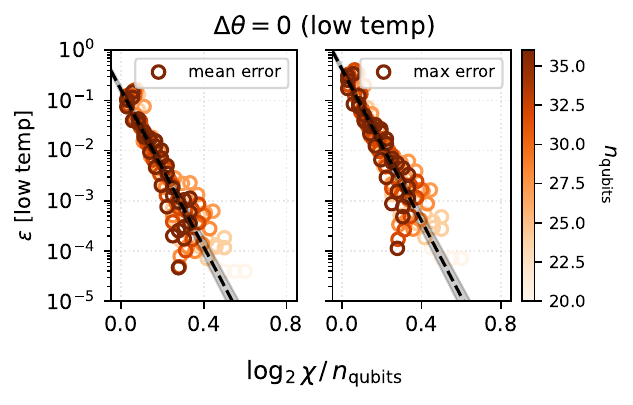}
    \includegraphics[width=0.48\textwidth]{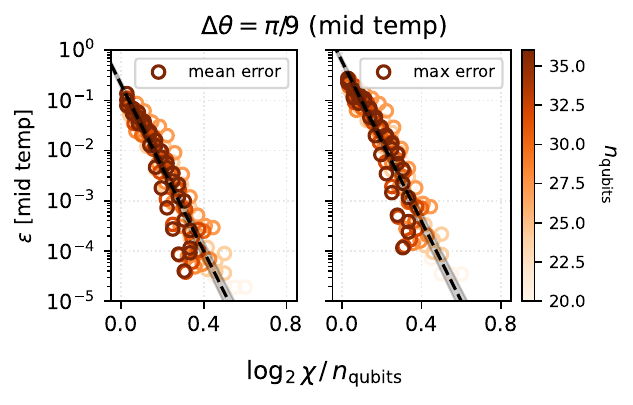}
    \includegraphics[width=0.48\textwidth]{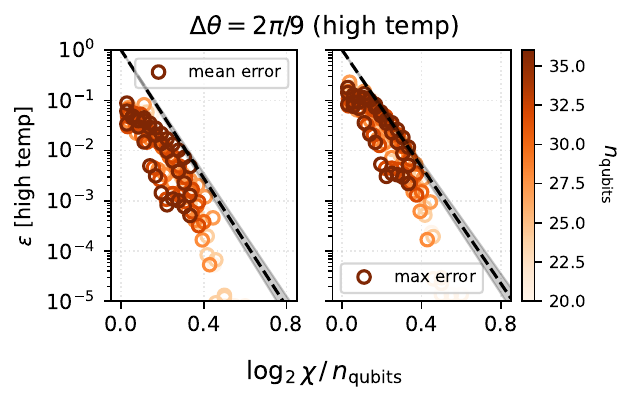}
    \caption{
      \textbf{Scaling of the absolute error between the rescaled
      $\boldsymbol{\Z{2}_{\gamma_{\rm fit}}}$ and the exact
      $\boldsymbol{\Z{2}_{\rm exact}}$ for the 3D TFIM, at fixed
      $\boldsymbol{(J,\, h,\, dt) = (-1, \,2, \,0.25)}$, across three different
      initial temperatures.} The plots show the mean absolute error (left) and
      the maximum absolute error (right) between $\Z{2}_{\gamma_{\rm fit}}$ and
      $\Z{2}_{\rm exact}$ over $s = 20$ steps for different grid sizes and bond
      dimensions $\chi$. The regression $\epsilon = K \, \chi^{\alpha/n}$ is
      computed considering only instances with at least $20$ qubits (see
      \Cref{tab:3d_scaling_error_params} for the fitting parameters). The shaded
      area corresponds to the linear regression error.
    }
    \label{fig:TFIM_3D_rescaling_error}
  \end{figure*}
}

\newcommand{\FigConvergenceLowMidHighTemp}{
  \begin{figure*}[h!]
    \centering
    \includegraphics[width=\textwidth]{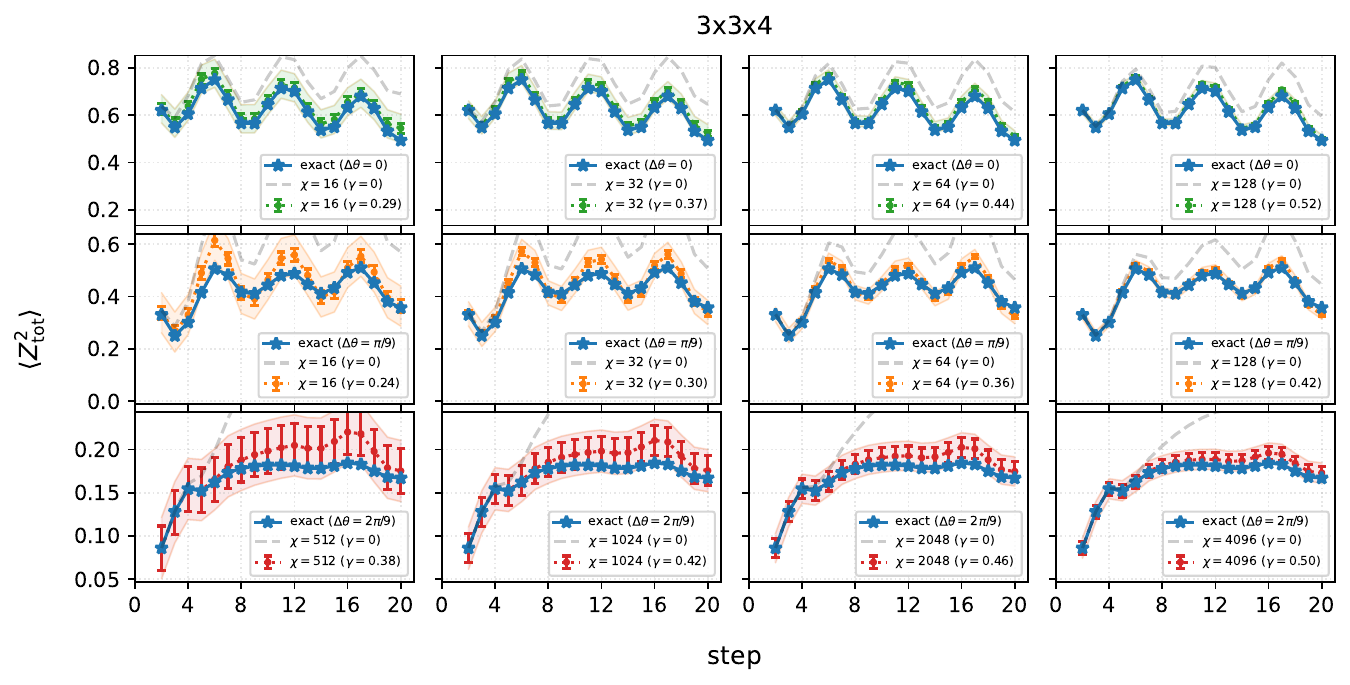}
    \caption{
      \textbf{Numerical simulations of a $\boldsymbol{3\times3\times4}$ TFIM for
        three different initial temperatures, with varying maximum bond
        dimension $\boldsymbol{\chi}$.} For each temperature, we identify the
      optimal $\gamma_{\rm fit}$ and the corresponding extrapolated error,
      following the procedures detailed for the mid-temperature case, as shown
      in \Cref{fig:TFIM_3D_gamma_scaling} and
      \Cref{fig:TFIM_3D_rescaling_error}. The blue-starred lines represent the
      exact results. The orange-circled lines correspond to our numerical
      results for $\Z{2}_{\gamma_{\rm fit}}$, with the error bars (shaded
      areas) corresponding to the estimates from the mean (maximum) absolute
      error from \Cref{fig:TFIM_3D_rescaling_error}. For comparison, the
      gray-dashed lines correspond to the value of $\Z{2}_{\rm MPS}$ without
      any rescaling.
    }
    \label{fig:TFIM_3D_conv_z2_low_mid_high_temp}
  \end{figure*}
}

\newcommand{\FigTFIMThreeDConvergence}{
  \begin{figure*}[h!]
    \centering
    \includegraphics[width=\textwidth]{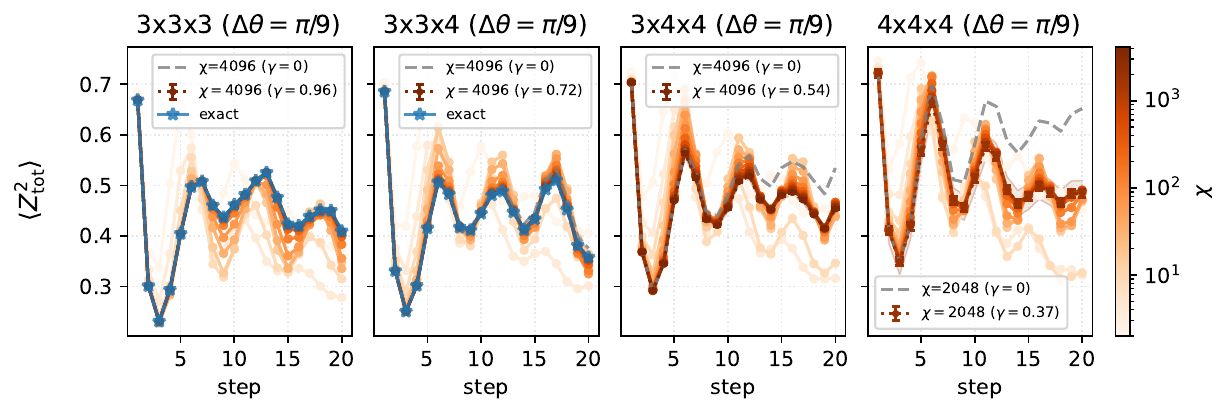}
    \caption{
      \textbf{(Left) Numerical simulations of different 3D TFIMs,
        with varying maximum bond dimension $\boldsymbol{\chi}$.} The
      blue-starred lines represent the exact results (when available). The
      orange-circled lines correspond to our numerical results for
      $\Z{2}_{\gamma_{\rm fit}}$, with the error bars (shaded areas)
      corresponding to the estimates of the mean (maximum) absolute error from
      \Cref{fig:TFIM_3D_rescaling_error}. The dashed lines correspond to the
      unrescaled $\Z{2}$.
    }
    \label{fig:TFIM_3D_conv_z2}
  \end{figure*}
}

\newcommand{\FigGammaScalingDensity}{
  \begin{figure*}[h!]
    \centering
    \includegraphics[width=0.8\textwidth]{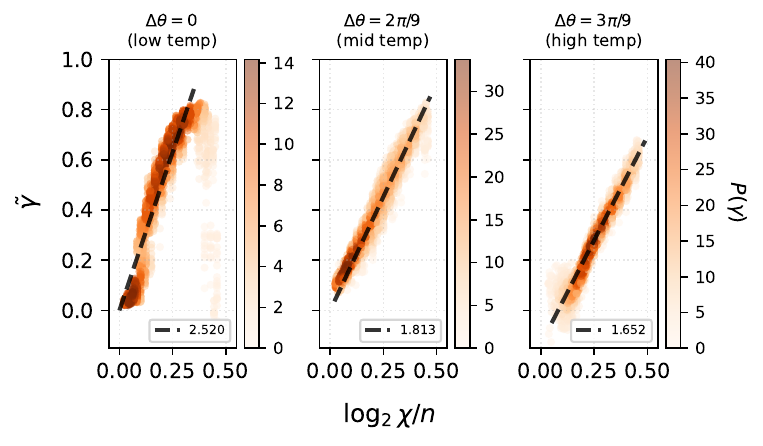}
    \caption{
      \textbf{Distribution of $\boldsymbol{\tilde\gamma}$ as a function of the
      rescaled bond dimension for three different initial temperatures of the 2D
      TFIM.} The panels show the probability distribution of $\tilde\gamma =
      \log\left( \Z{2}_{\rm ex} / \Z{2}_{\rm MPS} \right) / \log F_{\rm MPS}$
      obtained by varying the rescaled bond dimension $\log_2 \chi / n$ at a
      fixed initial temperature. Each point corresponds to a distinct data point
      for a specific system size and Trotter step $s$, colored according to the
      empirical probability distribution. To avoid bias from trivial data
      points, we only consider systems with at least $20$ qubits and Trotter
      steps $s \geq 10$.  The dashed black line corresponds to our extrapolated
      $\gamma^*$.
    }
    \label{fig:optimal_gamma}
  \end{figure*}
}

\newcommand{\FigXYInit}{
  \begin{figure*}[hbt]
    \centering
    \begin{tikzpicture}
      \node[anchor=south west,inner sep=0] (img1) at (0,0)
        {\includegraphics[width=0.32\textwidth]{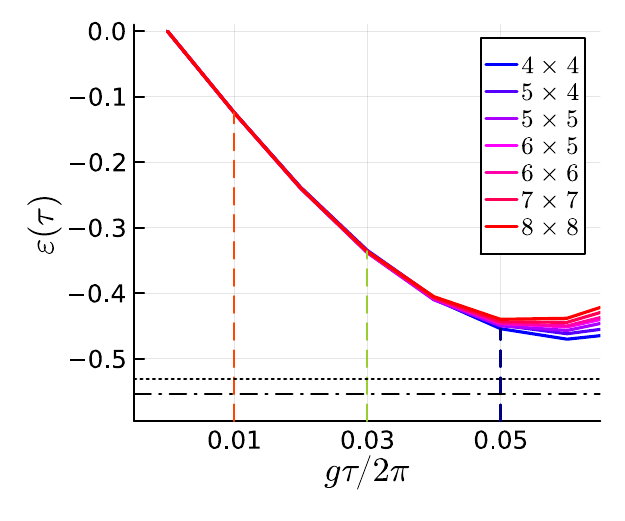}};
      \node[anchor=south west,xshift=0pt,yshift=2pt] at (img1.north west)
        {\textbf{(a)}};
    \end{tikzpicture}
    \begin{tikzpicture}
      \node[anchor=south west,inner sep=0] (img2) at (0,0)
        {\includegraphics[width=0.32\textwidth]{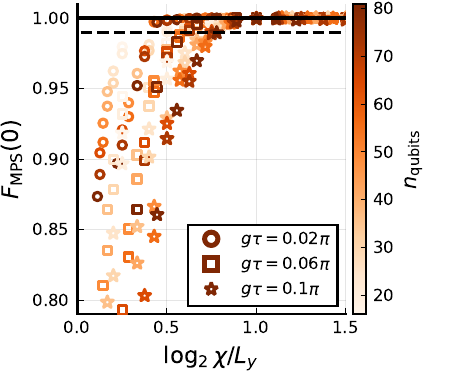}};
      \node[anchor=south west,xshift=0pt,yshift=2pt] at (img2.north west)
        {\textbf{(b)}};
    \end{tikzpicture}
    \begin{tikzpicture}
      \node[anchor=south west,inner sep=0] (img3) at (0,0)
        {\includegraphics[width=0.32\textwidth]{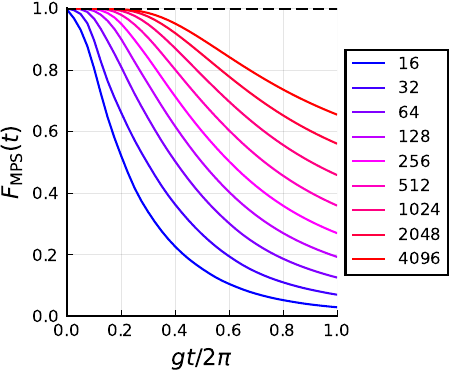}};
      \node[anchor=south west,xshift=0pt,yshift=2pt] at (img3.north west)
        {\textbf{(c)}};
    \end{tikzpicture}
    \caption{
      \textbf{${\rm U}(1)$-conserving initial states for XY quench dynamics.}
      \textbf{(a)} Energy per bond for the family of states described by
      \Cref{eq:xy_init} of the main text. Simulations in this supplement use $g \tau =
      0.10 \pi$, $g \tau = 0.06 \pi$, or $g \tau = 0.02 \pi$, which correspond
      to low-, intermediate-, and high-energy states, respectively (vertical
      lines from right to left). All values are higher than the predicted
      energy density of the BKT phase transition (dotted horizontal line),
      which is quite close to the ground-state energy density (dashed
      horizontal line). \textbf{(b)} MPS fidelity $F_{\rm MPS}(0)$ for
      preparing the initial state with MPS. In all cases, the initial state
      can be prepared with high fidelity using a bond dimension that scales
      exponentially only with the width. \textbf{(c)} MPS fidelity of the
      truncated time evolution, starting from the intermediate-energy state
      ($g \tau = 0.06 \pi$) on a $6 \times 6$ lattice for various bond
      dimensions $\chi$.
    }
      \label{fig:xy_init}
  \end{figure*}
}

\newcommand{\FigXYGammaScaling}{
  \begin{figure*}[t] 
    \centering

    \begin{minipage}[b]{0.86\textwidth}
      \centering
      \includegraphics[width=\textwidth]{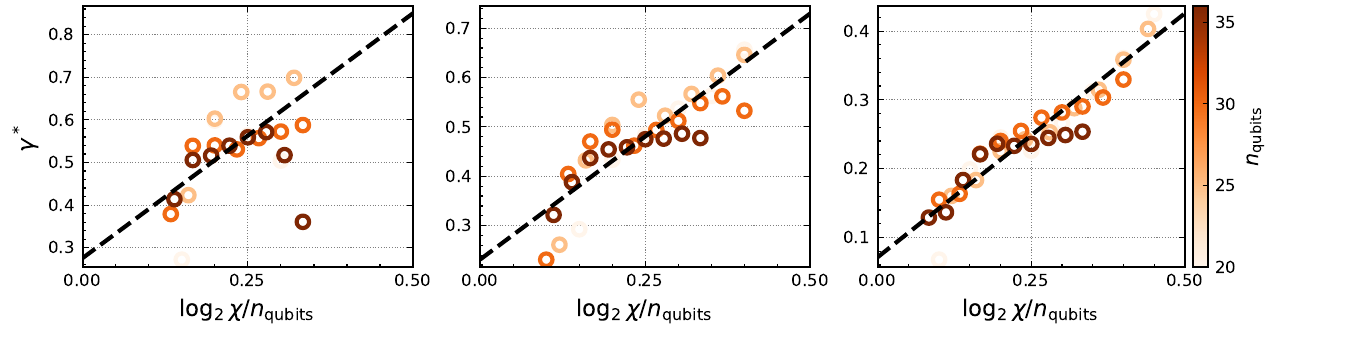}
    \end{minipage}
    \hfill
    \begin{minipage}[b]{0.13\textwidth}
      \raggedright
      \begin{tabular}{c|c|c}
        $\boldsymbol{g \tau}$ & $\boldsymbol{a}$ & $\boldsymbol{b}$ \\
        \hline \hline
        $0.10 \pi$ & $1.15$ & $0.28$\\
        $0.06 \pi$ & $1.00$ & $0.23$ \\
        $0.02 \pi$ & $0.71$ & $0.07$
      \end{tabular}
      \vspace{3em} 
    \end{minipage}

    \caption{
      \textbf{Fit for the $\gamma^*$ exponent and scaling parameters.} (Left)
      Correlation between the optimal exponent $\gamma^*$ and the rescaled bond
      dimension $\log_2\chi / n$. (Right) Fitting parameters for the linear
      model $\gamma_{\rm fit} = a \log_2(\chi)/n + b$. The fit is computed
      using systems containing between $20$ and $30$ qubits only, with the
      optimal values for the $n=36$ data shown for comparison. Although some
      deviations from the linear fit are evident, scaling with $\gamma$ taken
      from the linear fit still accelerates convergence substantially.
    }
    \label{fig:xy_gamma_scaling}
  \end{figure*}
}

\newcommand{\FigXYConvergence}{
  \begin{figure*}[htb]
    \centering
    \includegraphics[width=0.9\textwidth]{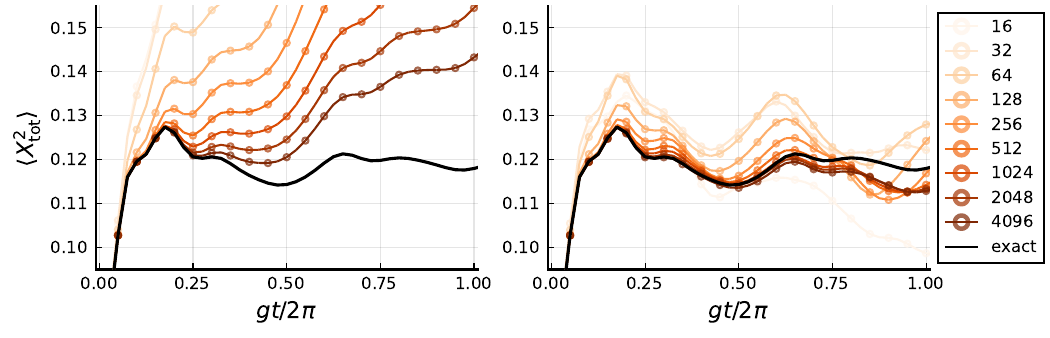}
    \caption{
      \textbf{Comparison of bond dimension convergence with and without
        rescaling.} The figures show the convergence of raw MPS data
      $\X{2}_{\rm MPS}$ (left) and rescaled MPS data
      $\X{2}_{\gamma_{\rm fit}}$ (right) with increasing bond dimension for the
      intermediate-energy state preparation on a $6 \times 6$ grid. The data
      converge toward the exact results at early times, while some deviation
      appears at late times. The convergence of the scaled data is accelerated
      compared to the unscaled data.
    }
    \label{fig:xy_convergence}
  \end{figure*}
}

\newcommand{\FigXYRescalingError}{
  \begin{figure*}[t!]
    \centering
    \begin{minipage}[b]{0.8\textwidth}
      \centering
      \includegraphics[width=\columnwidth]{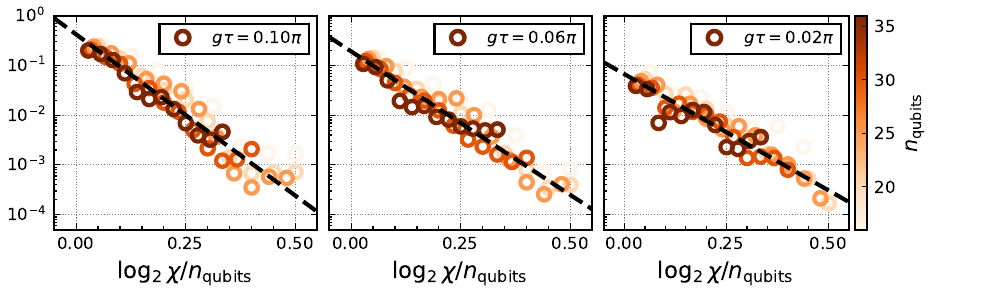}
    \end{minipage}
    \hspace{-5ex}
    \begin{minipage}[b]{0.10\textwidth}
      \raggedright
      \begin{tabular}{c|c|c}
        $\boldsymbol{g \tau}$ & $\boldsymbol{a}$ & $\boldsymbol{b}$ \\
        \hline \hline
        $0.10 \pi$ & $-6.5$ & $-0.4$\\
        $0.06 \pi$ & $-5.8$ & $-0.7$ \\
        $0.02 \pi$ & $-4.7$ & $-1.2$
      \end{tabular}
      \vspace{5em} 
    \end{minipage}
    \hfill
    \caption{
      \textbf{Scaling of the absolute error between the rescaled MPS and exact
        data.} {The plot shows the maximum absolute error between
        $\X{2}_{\gamma_{\rm fit}}$ and $\X{2}_{\rm exact}$ for the quench
        dynamics over the time range $0 \leq gt/2\pi \leq 1$, where the initial
        state is low energy (left), intermediate energy (middle), or high
        energy (right). In each case, most of the variation in the error size
        is described by the scaled bond dimension $\log_2 (\chi)/n$. Fits to
        $\log_{10} \epsilon = a \log_2 (\chi)/n + b$ are computed considering
        only instances with at least $20$ qubits, and the fitting parameters
        are reported in the table.}
    }
    \label{fig:xy_rescaling_error}
  \end{figure*}
}

\newcommand{\FigXYErrorValidation}{
  \begin{figure*}[t!]
    \centering
    \includegraphics[width=\textwidth]{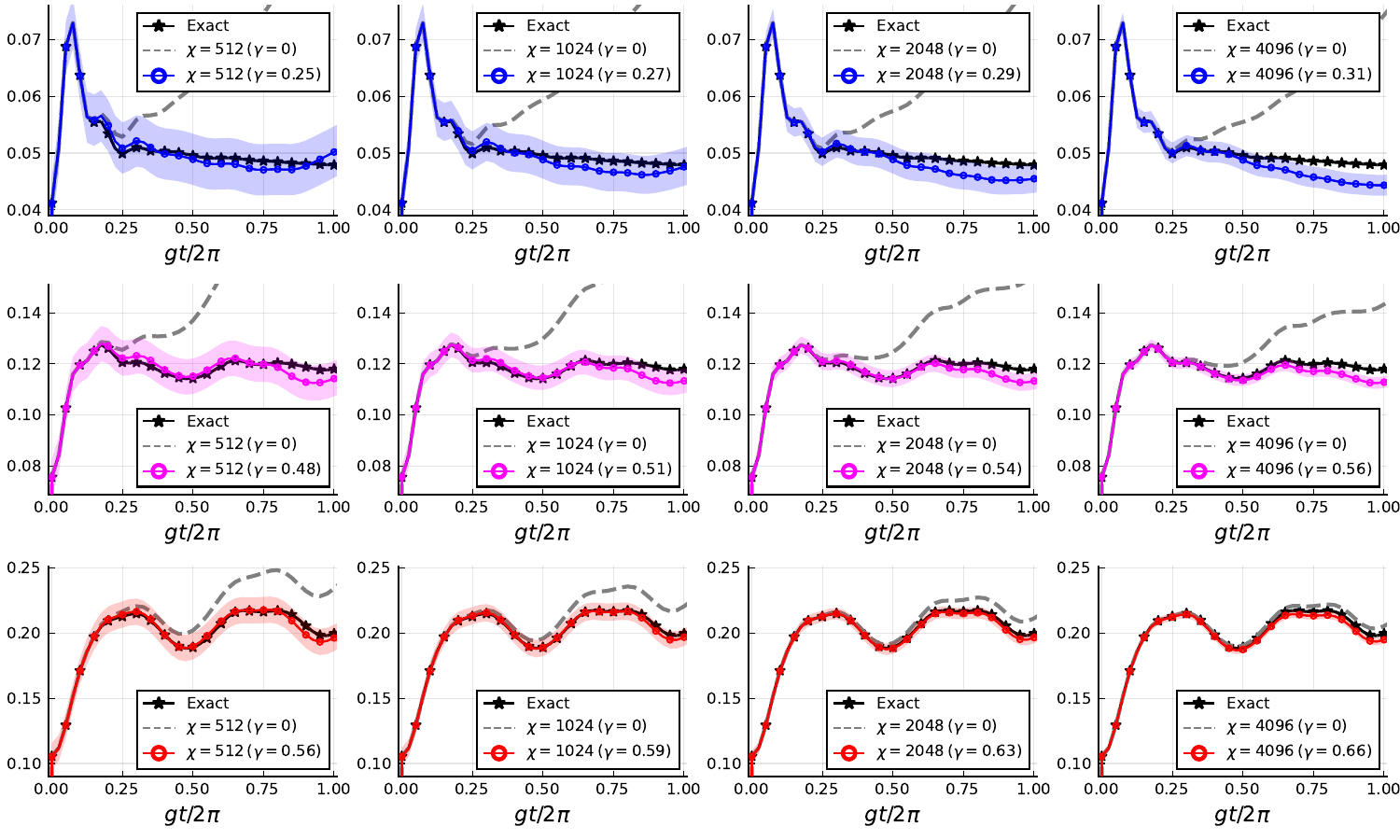}
    \caption{
      \textbf{Comparing predicted error with actual error in $6 \times 6$ XY
        quenches.} The figures show rescaled MPS data along with a predicted
      interval whose size is the error determined from the fits in
      \Cref{fig:xy_rescaling_error}, which estimates the maximum error across
      the full time window between the exact and rescaled MPS data.
      Additionally, exact results (black-starred lines) and unscaled MPS data
      (gray-dashed lines) are shown. We use the high-energy (top row),
      intermediate-energy (middle row), and low-energy (bottom row) initial
      states, respectively. In all cases, the exact data lie close to this
      interval.
    }
    \label{fig:xy_6x6_error_validation}
  \end{figure*}
}

\newcommand{\TableTFIMGamma}{
  \begin{table}[h!]
    \centering
    \begin{tabular}{c|c|c}
      $\boldsymbol{\Delta\theta}$ & $\boldsymbol{a}$ & $\boldsymbol{b}$ \\
      \hline
      \hline
      $0$ (low-temp)  & $2.654 \pm 0.006$ & $0$\\
      $\pi/9$ (mid-temp) & $2.153 \pm 0.001$ & $0$ \\
      $2\pi/9$ (high-temp) & $1.5104 \pm 0.0003$ & $0$
    \end{tabular}
    \caption{
      \textbf{Scaling of $\boldsymbol{\gamma_{\rm fit} = a \log_2(\chi)/n + b}$
        for the 3D TFIM, obtained by varying the bond dimension
        $\boldsymbol{\chi}$ and system size $\boldsymbol{n}$}. The scalings are
      reported in \Cref{fig:TFIM_3D_gamma_scaling}.
    }
    \label{tab:3d_gamma_scaling}
  \end{table}
}

\newcommand{\TableTFIMError}{
  \begin{table}[h!]
    \centering
    \begin{tabular}{c|c|c}
      $\boldsymbol{\Delta\theta}$ & $\boldsymbol{(K,\,\alpha)_{\rm mean}}$ &
      $\boldsymbol{(K,\,\alpha)_{\rm max}}$ \\
      \hline
      \hline
      $0$ (low-temp)  & $(0.16\pm0.03,\,-25.9\pm1.0)$ &
      $(0.41\pm0.07,\,-25.0\pm0.9)$ \\
      $\pi/9$ (mid-temp) & $(0.20\pm0.03,\,-27.5\pm1.0)$ &
      $(0.58\pm0.09,\,-26.4\pm0.9)$ \\
      $2\pi/9$ (high-temp) & $(1.00\pm0.00,\,-21.2\pm0.9)$ &
      $(1.00\pm0.00,\,-19.3\pm0.8)$
    \end{tabular}
    \caption{
      \textbf{Fitting parameters for the expected error $\boldsymbol{\epsilon =
        K\,\chi^{\alpha/n}}$ for the 3D TFIM, for different values of the
        initial parameter $\boldsymbol{\Delta\theta}$.}
      $(K,\,\alpha)_{\rm mean}$ and $(K,\,\alpha)_{\rm max}$ correspond to the
      fitting parameters for the expected mean and maximum absolute errors,
      respectively. For the high-temperature case, we fixed $K = 1$ to better
      capture the fluctuations.
    }
    \label{tab:3d_scaling_error_params}
  \end{table}

}

\newcommand{\TableSymbols}{
  \begin{table*}[h!]
    \centering
    \begin{tabular}{clr}
      \hline
      \multicolumn{3}{c}{\textbf{General Symbols}} \\
      \hline
      \textbf{Symbol} & \textbf{Description} & \textbf{First Appearance} \\[3pt]
      $H$ & Hamiltonian of the system & \Cref{eq:hamiltonian} \\
      $Z_i, X_i, Y_i$ & Pauli operators at site $i$ & \Cref{eq:hamiltonian} \\
      $s$ & Number of Trotter steps & \Cref{eq:z2_tot} \\
      $L_x, L_y, L_z$ & Lattice linear dimensions along each axis & \Cref{sec:tfim} \\
      $n$ & Total number of lattice sites/qubits ($n = L_x L_y$ or $n = L_x L_y L_z$) & \Cref{sec:tfim} \\
      $\chi$ & MPS bond dimension & \Cref{sec:results} \\
      $F_{\rm MPS}$ & Fidelity of the MPS wave function & \Cref{eq:Z_tot_via_F} \\
      $\gamma$ & Rescaling exponent & \Cref{eq:z2_tot_rescaled} \\
      $\gamma^*$ & Optimal rescaling exponent minimizing deviation from exact results & \Cref{sec:results} \\
      $\tilde\gamma$ & Exact rescaling exponent for individual simulations & \Cref{sec:results} \\
      $\epsilon$ & Maximum absolute error of the rescaled MPS observable & \Cref{sec:results} \\
      $t$ & Evolution time ($t = s dt$ or $t = s \delta t$) & \Cref{eq:xy_signal} \\
      & & \\
      \hline
      \multicolumn{3}{c}{\textbf{TFIM Symbols}} \\
      \hline
      \textbf{Symbol} & \textbf{Description} & \textbf{First Appearance} \\[3pt]
      $J$ & Spin-spin coupling constant for the TFIM & \Cref{eq:hamiltonian} \\
      $h$ & Transverse magnetic field strength & \Cref{eq:hamiltonian} \\
      $dt$ & Trotter step size for TFIM dynamics & \Cref{eq:trotter_step} \\
      $U(dt)$ & Unitary operator for a single Trotter step of the TFIM & \Cref{eq:trotter_step} \\
      $\theta$ & Initial state parameter for TFIM simulations & \Cref{eq:initial_state} \\
      $\ket{\Psi(\theta)}$ & Initial product state for TFIM dynamics & \Cref{eq:initial_state} \\
      $\theta_{\rm min}$ & State parameter minimizing the mean-field energy & \Cref{sec:results} \\
      $\Delta\theta$ & Deviation of initial state parameter from $\theta_{\rm min}$ & \Cref{sec:results} \\
      $\Z{}$, $\Z{2}$ & Magnetization and Ising order parameter expectation values & \Cref{eq:z2_tot} \\
      & & \\
      \hline
      \multicolumn{3}{c}{\textbf{XY Model Symbols}} \\
      \hline
      \textbf{Symbol} & \textbf{Description} & \textbf{First Appearance} \\[3pt]
      $H_{\rm XY}$ & Hamiltonian of the XY model & \Cref{eq:XY_hamiltonian} \\
      $g$ & Spin-spin coupling constant for the XY model & \Cref{eq:XY_hamiltonian} \\
      $\vec{X}_{\rm tot}$ & Total staggered magnetization vector of the XY model & \Cref{eq:xy_orderparam} \\
      $\xi$ & Correlation length of the XY model & \Cref{eq:xy_eq_corr} \\
      $T_{\mathrm{BKT}}$ & Berezenskii-Kosterlitz-Thouless transition temperature & \Cref{eq:xy_eq_corr} \\
      $\tau$ & Parameter tuning the initial state energy of the XY model & \Cref{eq:xy_init} \\
      $\ket{\psi_0(\tau)}$ & Initial state for XY quench dynamics & \Cref{eq:xy_init} \\
      $\ket{\rm Neel}$ & Néel product state & \Cref{eq:xy_init} \\
      $\delta\tau$ & Time step for the XY initial state preparation circuit & \Cref{sec:xy_stateprep} \\
      $\X{2}$ & Staggered magnetization correlation function of the XY model & \Cref{eq:xy_signal} \\
      $\delta t$ & Trotter step size for XY model dynamics & \Cref{eq:xy_trotter_step} \\
      $U(\delta t)$ & Unitary operator for a single Trotter step of the XY model & \Cref{eq:xy_trotter_step}
    \end{tabular}
    \caption{
      \textbf{Summary of main symbols and notation.}
    }
    \label{tab:symbols}
  \end{table*}
}

\begin{document}

\title{
  A Heuristic for Matrix Product State Simulation of Out-of-Equilibrium
  Dynamics of Two-Dimensional Quantum Spin Systems
}

\author{Salvatore Mandr\`a}
\email{smandra@google.com}
\affiliation{Google Quantum AI}
\author{Brayden Ware}
\affiliation{Google Quantum AI}
\author{Nikita Astrakhantsev}
\affiliation{Google Quantum AI}
\author{Sergei Isakov}
\affiliation{Google Quantum AI}
\author{Benjamin Villalonga}
\affiliation{Google Quantum AI}
\author{Tom Westerhout}
\affiliation{Google Quantum AI}
\author{Kostyantyn Kechedzhi}
\affiliation{Google Quantum AI}

\begin{abstract}
  Out-of-equilibrium dynamics of non-integrable Hamiltonian many-body quantum
  systems are characterized by highly entangled wave functions. Near-maximal
  entanglement arises in systems exhibiting thermalization or
  pre-thermalization, where the system converges to a steady state with a fixed
  energy density. Classical simulation of the time dependence of such wave
  functions requires exponential resources. However, typical computations aim to
  estimate expectation values of local operators and correlation functions to
  some expected precision. For thermalizing systems at sufficiently high energy
  densities, such computations can be done without storing the full wave
  function by instead simulating the evolution of the local operator, which
  requires significantly fewer resources.  Nonetheless, constructing such
  resource-efficient classical algorithms remains a challenge for intermediate
  energy densities, where simulating both the wave function and operator
  evolution is costly. In this paper, we propose a heuristic approach to
  accelerate the convergence of Matrix Product State (MPS) simulations of
  expectation values, applicable across a broad range of energy densities. We
  estimate the desired observables by rescaling the MPS results at low bond
  dimensions with a factor that depends on the fidelity of the MPS wave
  function. Using this technique, we simulated the dynamics of the
  two-dimensional Transverse-Field Ising Model (TFIM) on a $7\times8$ grid with
  periodic boundary conditions, using a maximum bond dimension of $\chi = 4096$
  on a single A100 GPU, as well as the dynamics of the two-dimensional XY model
  on grids of size up to $9\times9$. We compare our TFIM results to similar
  simulations on a digital quantum processor~\cite{Haghshenas2025Digital},
  demonstrating excellent agreement and confirming the predictive power of our
  method.
\end{abstract}

\maketitle

\section{Introduction}\label{sec:intro}

In recent years, rapid advancements in quantum hardware have revealed a
significant separation between the cost of realizing quantum sampling
algorithms on quantum processors and (classical) supercomputers
\cite{Arute2019Quantum, Morvan2024Phase, Andersen2025Thermalization,
Ransford2025Helios, Gao2025Establishing, Liu2025Robust}. These algorithms take
advantage of the ability of quantum processors to efficiently prepare and
process highly entangled states, whereas classical calculations require storing
and processing an exponential number of wave function coefficients to account
for entanglement.

This capacity to process highly entangled states also enables the efficient
simulation of quantum systems of interest in physics and chemistry in regimes
where the classical simulation of the entangled wave function is prohibitively
expensive. A number of practical applications of quantum processors rely on
quantum simulations of this type, where the task is to estimate the expectation
value of a local operator~\cite{Babbush2025Grand}. The least demanding
experiment is initialized in an easy-to-prepare state (e.g., a product state)
and subsequently evolves subject to the system's Hamiltonian; finally, the
expectation value of a local operator is measured.

For a system defined by a local Hamiltonian, estimating expectation values of
certain operators classically may require only sub-exponential resources.
Indeed, out-of-equilibrium dynamics initialized near the ground state of a
gapped Hamiltonian are often characterized by low (area law) entanglement and
allow for an efficient Matrix Product State (MPS) representation. Conversely,
initializing far from the ground state generates highly entangled dynamics.
However, this is often associated with the phenomenon of thermalization, where
expectation values of local observables can be computed by approximating the
system's state as locally thermal~\cite{Deutsch1991Quantum, Srednicki1994Chaos,
Deutsch2018Eigenstate}. At sufficiently high energy densities, the (equilibrium)
thermal state is described by a product operator~\cite{Bakshi2024High} and
therefore allows efficient computation of expectation values. In contrast,
out-of-equilibrium dynamics is significantly more complex.  Nonetheless, at
sufficiently high energy densities, correlation functions can be described by
efficient classical algorithms~\cite{Leviatan2017Quantum,
Rakovszky2022Dissipation}, whereas more complex observables present a
challenge~\cite{Mi2021Information, Abanin2025Constructive}. At intermediate
energy densities, constructing efficient classical algorithms even for
correlation functions has remained a challenge despite recent
progress~\cite{Begusic2025Real, Park2025Simulating}.

This intermediate density regime can be studied using quantum processors.
Quantum dynamics in the vicinity of a phase transition is of particular
interest, as it is characterized by high energy density and long-time
out-of-equilibrium dynamics. Such out-of-equilibrium dynamics have been studied
in a recent experiment~\cite{Haghshenas2025Digital} on the digital simulation
of the transverse field Ising model (TFIM), a prominent benchmark for quantum
magnetism and critical phenomena.

In this paper, we introduce a heuristic approach that accelerates the
convergence of classical MPS simulations. It allows computing expectation
values of local operators using a bond dimension significantly smaller than
that needed to represent the state with high precision. More precisely, our
heuristic involves rescaling the desired observables by a correction factor
that depends only on the total fidelity of the MPS, a parameter readily
available during the simulation.

We apply this technique to the dynamics of TFIMs with periodic boundary
conditions for grids up to $11\times11$. In particular, for the $7\times8$
TFIM, our numerical results agree with the experimental results reported in
Ref.~\cite{Haghshenas2025Digital} within the extrapolated numerical error,
confirming the predictive power of our numerical method.\\

\FigMPSDrawing

The paper is structured as follows: In \Cref{sec:tfim}, we introduce the TFIM
and the parameters used in our simulations; in \Cref{sec:results}, we provide
our numerical analysis and results for TFIMs up to $11\times11$, for which we
have exact numerical results for grids up to $6\times8$; in
\Cref{sec:quantinuum_h2}, we compare the available experimental data for the
$7\times8$ TFIM \cite{Haghshenas2025Digital} with our rescaling approach;
finally, in \Cref{sec:xy}, we apply our method to the 2D XY model. In the
Supplemental Material, we demonstrate the generalizability of our method to the
3D TFIM.

\section{Transverse-Field Ising Model}\label{sec:tfim}

For a general $D$-dimensional lattice, the TFIM
Hamiltonian is given by:
\begin{equation}\label{eq:hamiltonian}
  H = J \sum_{\langle i,j \rangle} Z_i Z_j + h \sum_i X_i = H_{ZZ} + H_{X}
\end{equation}
Here, $Z_i$ and $X_i$ are Pauli operators at site $i$, the sum $\sum_{\langle
i, j \rangle}$ runs over all pairs of nearest-neighbor sites with periodic
boundary conditions, $J$ is the coupling strength between neighboring spins,
and $h$ represents the strength of the transverse magnetic field.

\FigFiveBySixZ

It is well established that the model exhibits a quantum phase transition at
zero temperature, driven by the competition between the interaction term and
the transverse field~\cite{Onsager1944Crystal, Pfeuty1970One,
Sachdev1999Quantum, Schmitt2022Quantum}. In the case of a ferromagnetic
interaction ($J < 0$) and a small transverse field ($h \ll |J|$), the
interaction term dominates. This forces adjacent spins to align, causing the
system to spontaneously break the global $\mathbb{Z}_2$ symmetry. As the
strength of the transverse field $h$ increases, quantum fluctuations are
enhanced, tending to disorder the system by aligning the spins in the
$X$-direction. This competition leads to a quantum critical point at a finite
value of $|h / J|$, beyond which the long-range ferromagnetic order is
destroyed, and the system enters a quantum paramagnetic phase. Away from the
ground state in the pre-thermal regime, the TFIM also demonstrates
characteristics of a paramagnet-ferromagnet transition at finite temperature.

\FigZvsZZ

When simulating the dynamics of the TFIM on digital quantum computers, the
continuous time evolution is often approximated using a Suzuki-Trotter
decomposition. More precisely, the continuous dynamics of the TFIM is replaced
with a second-order Trotter expansion of the form:
\begin{equation}\label{eq:trotter_step}
  U(dt) \approx e^{-idt H_X/2} e^{-idt H_{ZZ}} e^{-idt H_X/2}.
\end{equation}
The unitary $U(dt)$ of a single Trotter step can then be realized by
interleaving layers of single-qubit gates with layers of two-qubit gates.

This method discretizes the evolution into a sequence of quantum gates,
effectively subjecting the system to a periodic drive. Such periodically
driven, or Floquet, many-body systems can exhibit a phenomenon known as Floquet
pre-thermalization. Instead of rapidly heating to an infinite-temperature state
as might be expected, the system can enter a long-lived, non-trivial steady
state that remains close to the initial state (in energy density) for a time
scale that is exponentially long in the driving frequency. This pre-thermal phase
is characterized by an emergent, nearly conserved effective Hamiltonian, which
governs the dynamics of the system before the eventual onset of thermalization.
As described in~\cite{Kuwahara2016Floquet, Abanin2017Effective,
Heyl2019Quantum, Ho2023Quantum}, such a pre-thermal phase can survive up to a
characteristic Floquet heating time scale on the order of $\tau_H \sim \exp(1 /
|J|dt)$. Therefore, the thermal physics of the TFIM can be accessed for an
exponentially long time scale $\tau_H$.\\

For our numerical simulations, we initialize the system in a product state of
the form:
\begin{equation}\label{eq:initial_state}
  |\Psi(\theta)\rangle = \bigotimes_j \Big(\cos(\theta/2)|0\rangle_j +
  \sin(\theta/2)|1\rangle_j\Big).
\end{equation}
On the pre-thermal timescale, the energy is approximately conserved and is
equal to its value in the initial state. The latter can be tuned in a broad
range by varying $\Delta\theta = \theta - \theta_{\rm min}$, where $\theta_{\rm
min} = \sin^{-1}(h/(dJ))$ corresponds to the minimum mean-field energy of the
TFIM, $H$ in \Cref{eq:hamiltonian}. Here $d$ is the number of neighbors of a
spin, on a square lattice $d=4$. For a sufficiently small $|h / J|$, it is
possible to identify the transition between the quantum ferromagnetic and
paramagnetic phases by studying the correlation function of the Ising order
parameter:
\begin{equation}\label{eq:z2_tot}
  \langle Z^{2}_{\rm tot}(s) \rangle = \frac{1}{n^2} \sum_{j,k} \langle
  \Psi(\theta) | (U^{\dagger})^{s} Z_{j} Z_{k} (U)^{s} | \Psi(\theta) \rangle,
\end{equation}
where $s$ corresponds to the total number of Trotter steps. Unless otherwise
specified, in this paper we focus on $L_x \times L_y$ grids with $dt = 0.25$, $J
= -1$, $h = 2$, and on three ``temperatures'': a low temperature ($\Delta\theta
= 0$), a mid temperature ($\Delta\theta = 2\pi/9$), and a high temperature
($\Delta\theta=3\pi/9$). The mid temperature is chosen to ensure that the
pre-thermal dynamics is close to the phase transition (see
\crtCrefname{figure}~2b of~\cite{Haghshenas2025Digital}). To simplify the
notation, we omit the explicit dependence on $s$ when clear from context.

\section{Numerical Results}\label{sec:results}

Our MPS simulations are performed using the Time-Evolving Block-Decimation
(TEBD) algorithm. This simulation evolves the initial state
$\left|\Psi(\theta_{\rm min} + \Delta\theta)\right\rangle$, as defined in
\Cref{eq:initial_state}, by sequentially applying the gates from the circuit
representation of the Trotter steps $(U(dt))^s$ in \Cref{eq:trotter_step}.

After each gate is applied, the MPS is returned to its canonical form, and a
truncation is performed via singular-value decomposition. As in standard TEBD,
the MPS is normalized to have unit norm after each truncation, while the
truncation fidelity is tracked separately. To ensure that the
computational cost of the singular-value decomposition depends only on the bond
dimension $\chi$, the orthogonality center is moved to the bond between the MPS
sites before performing the decomposition. To further minimize the computational
cost, multiple physical qubits are grouped into single MPS sites.  More
precisely, qubits in each row of an $L_x\times L_y$ grid are split into two
groups, and qubits in the same group are then fused together (see
\Cref{fig:mps_drawing}). When a two-qubit gate acts on qubits within the same
block, the corresponding tensor is updated locally, and the bond dimension
remains unchanged. In contrast, when a gate couples qubits in two distinct
blocks, it is decomposed as a Matrix Product Operator (MPO) and applied to the
MPS. For gates with a Schmidt rank of 2, such as $e^{-i\phi Z_i Z_j}$, this
operation effectively doubles the bond dimension for all virtual bonds located
between the two targeted blocks. This blocking technique improves the
efficiency of the TEBD algorithm by reducing the number of singular-value
decompositions required per time step. Crucially, because of the limited size of
the grouped sites, grouping sites together rather than using a one-to-one
mapping does not change our numerical results either qualitatively or
quantitatively (see \Cref{sec:ordering} of the Supplemental Material for
a discussion on ordering efficiency). For all simulations, unless otherwise
specified, the total number of Trotter steps is $s=20$, and the bond dimensions
are powers of $2$ ranging from $2$ to $4096$.  We performed all MPS simulations
on a single A100 GPU with $80\ \rm{GB}$ of RAM. To accommodate larger bond
dimensions, the MPS sites are stored in CPU RAM and transferred to GPU memory
only when required.  For the largest bond dimension of $\chi = 4096$ and $s =
20$ Trotter steps, the simulation took roughly 2 days.\\

\FigGammaScaling

It is instructive to first consider the expectation value of a single-body
observable:
\begin{equation}
  \langle Z_{\rm tot}(s)\rangle = \frac{1}{n} \sum_{j} \langle \Psi(\theta) |
  (U^{\dagger})^{s} Z_{j} (U)^{s} | \Psi(\theta) \rangle. \label{eq:Z_tot}
\end{equation}
Estimates of $\langle Z_{\rm tot}\rangle_{\rm MPS} = \bra{\psi_{\rm MPS}}
Z_{\rm tot}\ket{\psi_{\rm MPS}}$ for different bond dimensions, where
$|\psi_{\rm MPS}\rangle$ is the truncated wave function, are shown in
\Cref{fig:Z_tot_5x6}. We observe that the estimates at low bond dimensions
correlate with the exact results in the following sense. We define the fidelity
of the MPS wave function after $s$ time steps as $F_{\rm MPS}(s) =
\prod_{i=1}^s f_{\rm MPS}(i)$. Here, $f_{\rm MPS}(i)$ denotes the fidelity at
step $i$, calculated as the product of the truncation fidelities following each
gate application. For a theoretical analysis of the accuracy of such fidelity
estimates, see Ref.~\cite{zhou2020limits}. Then, we rescale the estimates for
$\Z{}$ for different bond
dimensions by $F_{\rm MPS}$, i.e., using the relation:
\begin{equation}
  \langle\psi|Z_{\rm tot}(s)|\psi\rangle \approx F_{\rm MPS}(s)
  \langle\psi_{\rm MPS} | Z_{\rm tot}(s) | \psi_{\rm MPS}\rangle.
  \label{eq:Z_tot_via_F}
\end{equation}
We observe that the estimates for different bond dimensions collapse
onto the same curve that approximates the exact result, as shown
in~\Cref{fig:Z_tot_5x6}.
It is worth contrasting our fidelity-based rescaling with standard extrapolation
techniques often used in tensor network simulations, such as Richardson
extrapolation~\cite{Marti2010DMRG}. Richardson extrapolation typically fits the
expectation value of an observable as a function of the discarded weight and
extrapolates to the zero-error limit, assuming a smooth functional dependence.
However, such an approach fails to accurately capture the dynamics in our case
(see also Fig.~3b of~\cite{Haghshenas2025Digital} for numerical results
obtained via zero-truncation extrapolation in MPS).\\

\FigConvergence

\FigRescalingError

This approximation can be understood from the following form of the MPS
decomposition of the wave function (after a single time step):
\begin{align}
  U\ket{\psi_{\rm MPS}(i-1)} &=
  \sqrt{f_{\rm MPS}(i)} \ket{\psi_{\rm MPS}(i)} \\\nonumber
  &\hspace{10px}+ \sqrt{1-f_{\rm MPS}(i)} \ket{\psi_{\perp}(i)},
\end{align}
where $\ket{\psi_{\rm MPS}(i)}$ and $|\psi_\perp(i)\rangle$ are, respectively,
the MPS wave function at step $i$ and its discarded part after truncation. By
construction, $\braket{\psi_{\rm MPS}(i) | \psi_\perp(i)}  = 0$. Note that for any
observable $O$, the orthogonality of the singular vectors implies
that $\bra{\psi_\perp(i)} O \ket{\psi_{\rm MPS}(i)}=0$. This property holds for
any single site observable or, more generally, any observable localized within
a region where the MPS has been returned to its canonical form. Therefore, the
approximation in \Cref{eq:Z_tot_via_F} amounts to neglecting the contribution
from $\ket{\psi_\perp(i)}$ to the final expectation value $\langle Z_{\rm
tot}\rangle$, under the assumption that $\bra{\psi_\perp} Z_{\rm tot}
\ket{\psi_\perp} \approx 0$.

\FigOrthogonalConvergence

Indeed, $F_{\rm MPS}$ corresponds to the weight of the retained wave function.
In our simulations, the expectation value estimate \Cref{eq:Z_tot_via_F}
has a relatively small error even when this weight is small, $F_{\rm MPS} < 1$.
In other words, only a small fraction of the norm of the wave function is
sufficient to approximate the expectation value. In contrast to the usual
estimate $\bra{\psi_{\rm MPS}} Z_{\rm tot}(s) \ket{\psi_{\rm MPS}}$,
\Cref{eq:Z_tot_via_F} relies on the properties of the observable. Indeed,
replacing $Z_{\rm tot}(s)$ with the identity operator yields a significant error
in \Cref{eq:Z_tot_via_F}.

\FigSixByEightLowMidHighTemp

This behavior can be intuitively understood in terms of out-of-equilibrium
properties of the two-dimensional TFIM, which demonstrates chaotic
characteristics~\cite{Mondaini2016Eigenstate, Mondaini2017EigenstateII}.
Indeed, the circuit $U$ in \Cref{eq:Z_tot} can be characterized by the spectrum
of the Floquet Hamiltonian, $U = \sum_E e^{-i E }\ket{E}\bra{E}$. Therefore, the
expectation value can be expanded in the eigenbasis,
\begin{gather}
  \langle O \rangle = \sum_{E,E^\prime} \braket{\Psi(\theta)|E}
  \braket{E^\prime|\Psi(\theta)} \braket{E|O|E^\prime} e^{-i(E-E^\prime)s}.\label{eq:O_vs_EEprime}
\end{gather}
The variance of the energy density in the initial state in
\Cref{eq:initial_state}, and therefore of the typical energy difference, scales
as $1/\sqrt{n}$, which is significant for the system sizes we consider.  At
sufficiently long times $s\gg1$, we expect the phases between sufficiently
distant energies to be approximately random.  Note that, in this limit, we
expect the average magnetization to vanish, as expected from the approximately
thermal statistics of local observables. The characteristic timescales for such
thermalization are short at high energies and much longer at the edge of the
spectrum.  The latter states typically have low entanglement and are therefore
captured by MPS with relatively low bond dimensions, justifying our observation
that \Cref{eq:Z_tot_via_F} holds even when the fidelity is low.

\FigLargeGrids

The case of the correlation function $\langle Z_{\rm tot}^2\rangle$ is more
complex. Ordered states can contribute an $\mathcal{O}(1)$ value to $\langle
Z^2_{\rm tot}\rangle$. Given the large transverse field in our setup ($h = 2$),
some of the ordered states are characterized by finite energy density and
volume-law entanglement, and contribute significantly to $\langle Z^2_{\rm
tot}\rangle$. Consequently, the discarded states $\ket{\psi_\perp}$ may have a
non-vanishing contribution, $\langle \psi_\perp|Z^2_{\rm tot}|\psi_\perp\rangle
\neq 0$. Nonetheless, we observe that low bond dimension estimates rescaled with
a fidelity-dependent factor correlate with the exact value. However, there is an
important difference compared to \Cref{eq:Z_tot_via_F}: the rescaling in
this case requires a different factor that is not simply the MPS fidelity
$F_{\rm MPS}$, as shown in \Cref{fig:z1_vs_z2},
\begin{equation}\label{eq:z2_tot_rescaled}
  \Z{2}_\gamma = \Z{2}_{\rm MPS} \cdot F_{\rm MPS}^{\gamma(n, \chi)}
\end{equation}
where the exponent $\gamma(n,\,\chi)$ is a function depending only on the
system size, $n = L_x\times L_y$, and the bond dimension, $\chi$. Note that, in
the limit $\chi \to \infty$, the MPS fidelity necessarily converges to unity,
$F_{\rm MPS}\to1$. Therefore, $\Z{2}_\gamma$ is also expected to converge to
the exact value $\Z{2}_{\rm exact}$ for a sufficiently large bond dimension.\\

To determine the scaling parameter $\gamma(n,\,\chi)$, we performed exact
simulations on grids up to $36$ qubits. We also performed exact simulations for
grids up to $6\times8$, leveraging the spatial and point-group symmetries
induced by the boundary conditions, to serve as a control test. The optimal
$\gamma^*$ is identified by minimizing the maximum absolute error among $s =
20$ Trotter steps between $\Z{2}_{\gamma^*}$ and $\Z{2}_{\rm exact}$. As shown
in \Cref{fig:gamma_scaling}, there is a clear correlation between the optimal
$\gamma^*$ and the rescaled bond dimension. The fitting parameters are reported
in \Cref{tab:gamma_scaling}. It is important to observe that this correlation
is independent of whether a one-to-one mapping or site grouping is used. While
the optimal scaling exponent could in principle assume arbitrary form, the
empirical distribution of the unconstrained exponent \mbox{$\tilde\gamma =
\log( \Z{2}_{\rm ex} / \Z{2}_{\rm MPS} ) / \log F_{\rm MPS}$}, defined as the
exponent that reduces the rescaling error to zero for each individual system
size and Trotter step is highly concentrated around the linear ansatz
$\gamma^*$ (see \Cref{fig:optimal_gamma} of the Supplemental Material).
\begin{table}[h!]
  \centering
  \begin{tabular}{c|c|c}
    $\boldsymbol{\Delta\theta}$ & $\boldsymbol{a}$ & $\boldsymbol{b}$ \\
    \hline
    \hline
    $0$ (low-temp)  & $2.588 \pm 0.003$ & $0$\\
    $2\pi/9$ (mid-temp) & $1.8252 \pm 0.0003$ & $0$ \\
    $3\pi/9$ (high-temp) & $1.641 \pm 0.001$ & $-0.1298 \pm 0.0001$
  \end{tabular}
  \caption{
    \textbf{Scaling of $\boldsymbol{\gamma_{\rm fit} = a \log_2(\chi)/n +
      b}$, by varying the bond dimension $\boldsymbol{\chi}$ and system size
    $\boldsymbol{n}$}. The scalings are reported in
    \Cref{fig:gamma_scaling}.
  }
  \label{tab:gamma_scaling}
\end{table}
\Cref{fig:convergence} shows the comparison between $\Z{2}_{\rm MPS}$ and
$\Z{2}_{\gamma_{\rm fit}}$ for a $5\times6$ TFIM with increasing bond
dimension. As seen in the figure, $\Z{2}_{\gamma_{\rm fit}}$ converges to
$\Z{2}_{\rm exact}$ much faster than the unrescaled $\Z{2}_{\rm MPS}$.
To provide a bound on the expected error for larger grids for which we do not
have exact results, in \Cref{fig:rescaling_error} we compare the mean absolute
error (left) and maximum absolute error (right) over $s = 20$ steps between
$\Z{2}_{\rm exact}$ and $\Z{2}_{\gamma_{\rm fit}}$ (the scaling for other
initial temperatures is reported in \Cref{fig:TFIM_rescaling_error} of
the Supplemental Material). More precisely, we observe
that the absolute error $\epsilon$ correlates strongly with the reduced bond
dimension $\log_2 \chi / n$, even for small $\chi$. This allows us to provide
an expected error estimate even for grids where exact results are
unavailable.\\

\begin{table}[h!]
  \centering
  \begin{tabular}{c|c|c}
    $\boldsymbol{\Delta\theta}$ & $\boldsymbol{(K,\,\alpha)_{\rm mean}}$ &
    $\boldsymbol{(K,\,\alpha)_{\rm max}}$ \\
    \hline
    \hline
    $0$ (low-temp)  & $(0.18\pm0.02,$ & $(0.40\pm0.04,$ \\
    & $-25.5\pm0.7)$ & $-24.8\pm0.6)$ \\
    $2\pi/9$ (mid-temp) & $(0.24 \pm 0.02,$ & $(0.59 \pm0.06,$ \\
    & $-24.4 \pm 0.6)$ & $-22.6 \pm 0.6)$ \\
    $3\pi/9$ (high-temp) & $(0.46\pm0.05,$ & $(1.3\pm0.2,$ \\
    & $-24.2\pm0.7)$ & $-24.0\pm0.8)$
  \end{tabular}
  \caption{
    \textbf{Fitting parameters for the expected error $\boldsymbol{\epsilon
      = K\,\chi^{\alpha/n}}$, by varying the initial parameter
    $\boldsymbol{\Delta\theta}$.} $(K,\,\alpha)_{\rm mean}$ and
    $(K,\,\alpha)_{\rm max}$ correspond to the fitting parameters for the
    expected mean absolute error and the expected maximum absolute error
    respectively.
  }
  \label{tab:scaling_error_params}
\end{table}

In
\Cref{fig:orthogonal} we compare the error of our rescaling method against the
standard unrescaled estimate and a more complete approximation that explicitly
includes the discarded component $|\psi_\perp\rangle$. Our results demonstrate
that the off-diagonal terms $\langle \psi_\perp | Z^2_{\rm tot} | \psi_{\rm
MPS}\rangle$ have a negligible contribution, and that the rescaled estimate is
significantly more accurate than the unrescaled one and performs comparably to
the approximation that includes $|\psi_\perp\rangle$. Note that the vanishing of the
off-diagonal terms is consistent with the qualitative picture of gradual
randomization of phases in \Cref{eq:O_vs_EEprime} that we introduced above.
\Cref{eq:z2_tot_rescaled} suggests that additional MPS truncation error
compared to \Cref{eq:Z_tot_via_F} has an effect of an overall damping
factor. This invites an analogy with a white noise error model in random
circuits~\cite{Arute2019Quantum}, where errors translate into a damping factor
upon random twirling of the state of the system. Within the scope of this
paper, \Cref{eq:z2_tot_rescaled} is an empirical observation. We leave further
theoretical justification of this property for future study.

\FigSevenByEightZZ

In \Cref{fig:6x8_z2_low_mid_high_temp}, we compare our rescaling technique
against the exact results for the $6\times8$ TFIM across the three different
effective temperatures. For each case, we identified the optimal $\gamma_{\rm
fit}$ and the corresponding extrapolated error, following the procedures
detailed for the mid-temperature case, as shown in \Cref{fig:gamma_scaling} and
\Cref{fig:rescaling_error}. The blue-starred lines represent the exact results,
while the orange-circled lines show our numerical results for
$\Z{2}_{\gamma_{\rm fit}}$. The error bars (shaded areas) correspond to the
estimated mean (maximum) absolute errors. In all instances, our rescaling
approach successfully predicts the exact value of $\Z{2}$ within the estimated
maximum error bound.

We have also extended our simulations to larger TFIM grids up to $11\times11$,
with a maximum bond dimension of $\chi = 4096$ (and $\chi = 2048$ for the
$11\times11$ grid). Due to memory limitations, we used a one-to-one mapping for
these simulations instead of grouping sites. The results are summarized in
\Cref{fig:large_grids}, which shows the convergence of $\Z{2}_{\gamma_{\rm
fit}}$ with increasing bond dimension $\chi$ for different grid sizes. The
purple-shaded area and the orange-shaded areas correspond, respectively, to the
mean absolute error and the maximum absolute error extrapolated from
\Cref{fig:rescaling_error} for the largest available bond dimension for the
given grid size. Unfortunately, without verification against exact or
error-bounded simulations, we are unable to guarantee that the results we
observe in \Cref{fig:large_grids} are converging to the correct $\Z{2}$ for the
largest systems. A better understanding of the functional form of the factor and
a more rigorous analysis of the error of this approximation may allow computing
$\Z{2}$ without relying on any extrapolation. We leave this for future study.\\

\section{Comparison to Experimental Results on
\texorpdfstring{$\boldsymbol{7\times8}$}{7x8} grids}\label{sec:quantinuum_h2}

In a recent paper~\cite{Haghshenas2025Digital}, the TFIM was
simulated on a 56-qubit trapped-ion quantum processor, the Quantinuum H2. This
gate-based machine is universal, allowing for the flexible implementation of
quantum simulations. The native two-qubit gate implemented on the hardware is
the $U_{ZZ}(\phi) = \exp({-i\phi Z \otimes Z /2})$ gate, which is used directly
to construct the Trotter steps for the simulation without further
decomposition.

Using a combination of different error mitigation protocols, the authors of
Ref.~\cite{Haghshenas2025Digital} simulated the quantum dynamics of the
$7\times8$ TFIM up to $s = 20$ Trotter steps (see \crtCrefname{figure}~3b
in~\cite{Haghshenas2025Digital}). To benchmark their quantum simulations, the
authors employed a comprehensive suite of state-of-the-art classical numerical
methods, including projected entangled pair state (PEPS) simulations with
various compression schemes such as belief propagation. However, these
classical methods fail after just a few Trotter
steps~\cite{Haghshenas2025Digital}.

\FigThermal

In contrast, we demonstrate that the correlation function of the Ising order
parameter $\Z{2}$ can be classically simulated for the $7\times8$ TFIM using a
bond dimension of $\chi = 4096$.
\Cref{fig:7x8_z2} (Left) shows our numerical simulations for the $7\times8$
TFIM with varying maximum bond dimensions. The blue-starred and blue-triangled
lines represent the results from the Quantinuum H2 device
before~\cite{Haghshenas2025Digital_v1} and after~\cite{Haghshenas2025Digital}
the improved error mitigation protocol, respectively. The orange-circled lines
correspond to our numerical results for $\Z{2}_{\gamma_{\rm fit}}$, with the
error bars (shaded areas) corresponding to the estimates from the mean
(maximum) absolute error from \Cref{fig:rescaling_error}. More precisely, our
numerical results are in excellent agreement with the new experimental results
from the Quantinuum H2 device, which are now within our estimated error bounds.
The accuracy of our numerical results is further corroborated by the fact that
our MPS prediction $\Z{2}_{\gamma_{\rm fit}}$ oscillates around the expected
thermal value for the Floquet Hamiltonian (to $\mathcal{O}(dt^2)$), computed in
the canonical ensemble via an MPS purification
ansatz~\cite{Haghshenas2025Digital}, as shown in \Cref{fig:thermal}. Finally,
\Cref{fig:7x8_z2} (Right) shows the comparison of both the MPS prediction and
experimental results against exact results obtained using exact tensor network
contraction, up to $s=4$.

\section{XY Model}\label{sec:xy}

To evaluate the generalizability of the rescaling heuristic beyond the
transverse-field Ising model (TFIM), we apply it to the quench dynamics of the
two-dimensional XY model. This choice introduces several physical ingredients
absent from the TFIM dynamics considered in the previous sections a conserved
$\mathrm{U}(1)$ charge in the system and the presence of vortex excitations in
addition to spin flip (magnon) excitations which change the type of order in
the system.

The XY Hamiltonian is given by:
\begin{equation}\label{eq:XY_hamiltonian}
  H_{XY} = \frac{g}{2} \sum_{\langle i,j \rangle} \left(X_i X_j + Y_i Y_j \right)
\end{equation}
Here, $X_i$ and $Y_i$ are Pauli operators at site $i$, the sum runs over all
pairs of nearest-neighbor sites, and $g$ is the coupling strength between
neighboring spins. In this section, we consider open boundary conditions on
two-dimensional rectangular grids of size $L_x \times L_y$, with $n=L_x L_y$
total sites.

The XY model exhibits a finite-temperature phase transition of the
Berezenskii-Kosterlitz-Thouless type~\cite{Berezinskii1972Destruction,
Kosterlitz1972Long, Kosterlitz1973Ordering}. The equilibrium properties of this
model, including the properties of its critical point, have been
characterized with quantum Monte Carlo simulations~\cite{Loh1985Monte,
Ding1990Kosterlitz, Ding1992Phase, Zhang1992Greens, Olsson1995Monte}. The order
parameter sensitive to the phase transition is the staggered, in-plane
magnetization
\begin{equation}
  \vec{X}_{\mathrm{tot}} = \frac{1}{n}\sum_j (-1)^j \vec{X}_j,
  \label{eq:xy_orderparam}
\end{equation}
where $\vec{X}_j$ denotes the two-component vector $\{X_j, Y_j\}$, and
$(-1)^j$ represents a staggered pattern of $\pm1$ for lattice sites $j$. For
an equilibrium thermal state $\exp (-\beta H_{\mathrm{XY}})$, the phase
transition is detectable via correlations of the order parameter components,
which take the form:
\begin{equation}
  \left \langle (-1)^r X_i X_{i + r}\right \rangle
  \sim r^{-\lambda} e^{-r / \xi},
  \label{eq:xy_eq_corr}
\end{equation}
where $\xi$ diverges at and below the critical temperature
$T_{\mathrm{BKT}}$\cite{Ding1990Kosterlitz}. The dynamics of the
order parameter, both equilibrium and the non-equilibrium quench studied
here, have not been thoroughly explored, partly due to the challenge of
classically simulating two-dimensional quantum dynamics at finite energy
density.\\

\subsection{State Preparation and Dynamics}\label{sec:xy_stateprep}

We aim to prepare initial states with definite values of the conserved
$\mathrm{U}(1)$ charge across a range of distinct energy densities. Product
states are insufficient for this purpose: product states where each spin aligns
with the $Z$-axis have definite values of $\Z{}$, but $\langle H_{\rm XY}
\rangle = 0$ for all such states. Conversely, product states of spins rotated
away from the $Z$-axis do not have definite values of $\Z{}$.
Instead, we use the following family of initial states to begin
our quench dynamics:
\begin{equation}
  \ket{\psi_0(\tau)} = \exp(-i \frac{\pi}{8} H_Z )
  \exp(- i \tau H_{\mathrm{XY}} ) \ket{\mathrm{Neel}},
  \label{eq:xy_init}
\end{equation}
where $H_Z = \sum_j (-1)^j Z_j$, and \(\ket{\mathrm{Neel}}\) is the ground
state of $H_Z$. These states are notable because they access a relatively wide
range of energies with small values of $\tau$ (see \Cref{fig:xy_init}(a) of
the Supplemental Material). Since $\tau$ is small, \(\ket{\psi_0(\tau)}\) can
be prepared with a shallow digital quantum circuit after Trotterization of the
$H_{\rm XY}$ evolution as described in \Cref{sec:xy}.
More importantly for this work, these initial states can be easily prepared as
MPS without incurring significant truncation errors for the system sizes
considered. Specifically, we focus on three states with values $g \tau = 0.02
\pi$, $g \tau = 0.06 \pi$, and $g \tau = 0.10 \pi$, which cover a broad range
of initial energy densities (see \Cref{fig:xy_init}(a) of the Supplemental
Material). We use a time step of size $g \delta \tau = 0.02 \pi$ for the
numerical simulations of this state preparation, which correspond to $1$, $3$,
and $5$ Trotter steps, respectively. The evolution for this state preparation
is sufficiently short—and the generated entanglement sufficiently small—that
these states are captured with high accuracy using MPS of modest bond
dimensions. As reported in \Cref{fig:xy_init}(b) of the Supplemental
Material, the necessary bond dimension scales exponentially with the width
$L_y$ of the rectangular grid, the expected scaling for states with only
short-range entanglement.\\

Unlike the Ising model simulations, the states we consider are all in the
high-temperature phase of the XY model. The energy density of the BKT
transition is known from Ref.~\cite{Ding1992Phase}, which reports a value of
$\epsilon_c \approx -0.53g$ for the energy density per bond near the transition
temperature. This value is not reached by the family of states in
\Cref{eq:xy_init}. Incidentally, this energy density is quite close to the
ground-state energy density $\epsilon_g \approx -0.55g$. Consequently, for the
sizes $n$ considered in this paper, states with energy density less than
$\epsilon_c$ have very few excitations above the ground state.\\

The signal we consider is the dynamics of the total correlation function
\begin{equation}
  \X{2}(t) = \frac{1}{n^2} \sum_{j,k} (-1)^{j-k}
  \left \langle X_j X_k \right \rangle(t),
  \label{eq:xy_signal}
\end{equation}
which can be expressed as a second moment of a component of the order parameter
\Cref{eq:xy_orderparam}. As we are in the high temperature phase, we expect via
\Cref{eq:xy_eq_corr} that in equilibrium this quantity decays like $1/n$ for
large $n$.\\

For the purposes of our dynamics simulations, we approximate the continuous time
evolution of the XY Hamiltonian with a Suzuki-Trotter decomposition.  We use the
depth 7, second-order decomposition based on splitting the Hamiltonian into $4$
sets of terms that each act only on non-overlapping sets of bonds.
\begin{align}\label{eq:xy_trotter_step}
  U(\delta t) \approx & e^{-i\delta t H_A/2}  e^{-i \delta t H_B/2}
  e^{-i\delta t H_C/2} e^{-i \delta t H_D} \cdot \nonumber \\
  & e^{-i\delta t H_C/2} e^{-i\delta t H_B/2} e^{-i\delta t H_A/2}.
\end{align}
Specifically, this splitting divides the bonds into horizontal bonds starting on
even ($H_A$) or odd ($H_B$) columns and vertical bonds starting on even ($H_C$)
or odd ($H_D$) rows. The Trotter step size $\delta t$ is chosen to ensure that
the contributions of the Trotter error are negligible for the purposes of our
MPS error analysis; we find that setting $g \delta t = 0.05 \pi$ is sufficient
for this purpose. Our quench dynamics analysis focuses on the time range $0
\leq gt \leq 2 \pi$, for which the total number of Trotter steps is $40$.\\

\subsection{Numerical Results}\label{sec:xy_numerics}

\FigXYComparison

As described in \Cref{sec:results}, our numerical simulations are performed
with the Time-Evolving Block-Decimation (TEBD)
algorithm~\cite{Vidal2004Efficient}. The simulation first constructs the
initial state \Cref{eq:xy_init} as a MPS subsequently evolves it by
sequentially applying $U(\delta t)$ defined in \Cref{eq:xy_trotter_step}. In
contrast to the TFIM simulations, for the XY model we use a one-to-one mapping
of qubits to MPS sites and do not block multiple physical qubits into single
MPS sites.

Our exact simulations of the XY model cover rectangular grids of sizes $4
\times 4$, $5 \times 4$, $5 \times 5$, $6 \times 5$, and $6 \times 6$; MPS
simulations are done for these grids with bond dimensions up to $4096.$
Throughout, we use fits derived by comparing MPS to exact data only on grids up
to $6 \times 5$ qubits, reserving the exact $6 \times 6$ data for validating
the derived fits. Additionally, we show results from MPS simulations on grids
of size $7 \times 6$, $7 \times 7$, $8 \times 8$, and $9 \times 9$ to
demonstrate how our extrapolations yield predictions for dynamics beyond the
reach of exact simulation. As for the TFIM, the total truncated weight in our
MPS simulations is tracked through the MPS fidelity $F_{\mathrm{MPS}}$ (see
\Cref{fig:xy_init}(c) of the Supplemental Material).

Our analysis of the order parameter dynamics proceeds analogously to the TFIM
case. For each initial state, bond dimension $\chi$, and system size $n$, we
determine the optimal $\gamma^*$ that minimizes the distance between the exact
signal $\X{2}_{\rm exact}$ and the rescaled MPS signal $\X{2}_{\rm MPS}
F^{\gamma}_{\rm MPS}$ across the considered time range (reported in
\Cref{fig:xy_gamma_scaling} of the Supplemental Material). As for the TFIM,
we observe a correlation between $\gamma^*$ and the rescaled bond dimension
$\log_2 \chi/ n$, albeit with some significant deviations. We will see below
that despite these deviations, rescaling by $F_{\rm MPS}^{\gamma}$ using a
simple linear model for $\gamma$ still greatly accelerates convergence on
system sizes both within and beyond the reach of our training data set,
compared to numerics that are unscaled or scaled with $F_{\rm MPS}$. We
introduce two deviations from the fitting procedure used for the TFIM. First,
we find that bond dimensions that are too small to accurately represent the
initial state $\ket{\psi_0(\tau)}$ show distinct trends in $\gamma^*$ and
poor-quality dynamics $\X{2}_{\rm MPS}(t).$ Thus, we exclude these bond
dimensions from our fits and subsequent plots; specifically, we use the cutoff
$F_{\rm MPS}(t=0) > 0.99$ (see \Cref{fig:xy_init}(b) of the Supplemental
Material). Second, we observe that for small sizes and very large bond
dimensions, where the fidelity remains close to $1$ across all times,
$\X{2}_{\rm MPS} F^{\gamma}_{\rm MPS}$ is highly insensitive to the choice of
$\gamma$. In these cases, the resulting $\gamma^*$ does not exhibit any
discernible trend. We use a cutoff $F_{\rm MPS}(t_f)<0.95$ to ensure
sensitivity as $\gamma$ is varied, where $t_f$ is the final time included in
the fit.

To see how rescaling by $F_{\rm MPS}^{\gamma}$ performs, we first plot a
comparison of rescaled MPS and corresponding exact data for $6 \times 6$
systems with a bond dimension of $4096$ in \Cref{fig:xy_6x6_4096}. The $6
\times 6$ exact data was not used to predict the exponent $\gamma$ used here;
instead, the rescaling exponent is drawn from the fits reported in
\Cref{fig:xy_gamma_scaling} of the Supplemental Material, which use data up
to $30$ qubits only. We find that the accuracy of the rescaled data is quite
high for early times but begins to deviate gradually as time progresses. We
show $40$ Trotter steps here, each of size $g \delta t = 0.05 \pi$, compared to
the $20$ Trotter steps shown in \Cref{sec:results}; our accuracy is higher when
restricted to only $20$ Trotter steps.

Next, we plot the same for a wide range of bond dimensions to investigate the
convergence of the rescaled data (see \Cref{fig:xy_convergence} of the
Supplemental Material for intermediate temperatures on the $6 \times 6$ grid).
We find that convergence to the exact data is greatly accelerated by rescaling
compared to the unrescaled data. In both cases, the error tends to increase
with the evolution time.

To characterize the error of the rescaled data, we plot the maximum error
between the rescaled and exact data over the time range $0 \leq gt / 2 \pi \leq
1$ for all system sizes and bond dimensions for which we have exact data. As
for the TFIM, the scaling of the error $\epsilon$ is well described by a linear
relation between $\log \epsilon$ and the scaled bond dimension $\log_2 \chi /
n$. These fits are consistent with the behavior observed for the $6 \times 6$
grid data across a range of bond dimensions (see
\Cref{fig:xy_6x6_error_validation} of the Supplemental Material), giving us
confidence that the error predictions will also apply to larger sizes. We note
that while the exact data lies slightly outside the plotted intervals in some
cases, this is expected because the error interval plots the typical value of
the error (maximized over the time window; see \Cref{fig:xy_rescaling_error}
of the Supplemental Material) rather than an upper bound.

\FigXYLargeGrids

Finally, we use our rescaling and error fits to extrapolate predictions beyond
the reach of exact numerics. The results are shown in \Cref{fig:xy_large_grids}.
Since the error estimate predicts a maximum error over the full time window,
whereas actual errors are typically much smaller at early times, the
convergence of the rescaled MPS is expected to be better than predicted for
early times. As the extrapolation is extended to larger sizes, our error
estimates become large compared to the physical signal.

\section{Conclusions}\label{sec:conclusions}

In this paper, we propose a heuristic approach to accelerate the convergence of
Matrix Product State (MPS) simulations, applied to the two-dimensional quantum
spin systems. We observe that a smaller bond dimension than previously expected
may be sufficient to simulate physical observables. This is because, for some
observables, a low bond dimension estimate is already correlated with the exact
value, differing from it to a good approximation only by a rescaling factor.
More specifically, in the case of transverse field Ising model the
magnetization $\Z{}$, can be estimated by rescaling a low bond dimension MPS
estimate $\Z{}_{\rm MPS}$ by the fidelty of MPS $F_{\rm MPS}$. In the case of
correlation function of the Ising order parameter $\Z{2}$, rescaling by $F_{\rm
MPS}$ alone is insufficient. However, we empirically observe that using a
scaling factor of the form $F_{\rm MPS}^{\gamma(n,\,\chi)}$ is sufficient to
accelerate the convergence of $\Z{2}$ from numerical MPS simulations. The
parameter $\gamma(n,\,\chi)$ is extrapolated from numerical simulations on
grids up to $36$ qubits and extended to larger grids. For sizes where exact
numerical simulations are not available, it is instructive to compare our
results to the available experimental data—specifically, the recently reported
$7\times8$ TFIM. Not only were we able to accurately predict the converged
results for $\Z{2}$ prior to the release of Quantinuum's improved error
mitigation~\cite{Haghshenas2025Digital_v1}, but the updated experimental
results~\cite{Haghshenas2025Digital} also match our numerical results within
the estimated error bounds, demonstrating the predictive power of our method.
We also extended our simulations to larger TFIMs up to $11\times11$.  However,
our extrapolation is based on exact results for systems of limited size, and it
might not be reliable for such large systems. We find that this behavior
generalizes to other quantum spin systems. More specifically, we observe a
similar accelerated convergence of MPS in simulations of the quench dynamics of
the 2D XY model.

\section{Contributions}

S.M. designed the rescaling heuristic, developed the TFIM MPS simulator, and
ran the 2D and 3D TFIM simulations. B.W. developed the XY model MPS simulator
and ran the 2D XY simulations. K.K. developed the theoretical analysis and
interpretation. N.A. and T.W. designed the exact simulator and ran exact
simulations for grids up to $6\times7$. S.I. designed an alternative exact
simulator and ran exact simulations for grids up to $6\times8$. B.V. designed
the tensor network simulator and ran exact simulations on $7\times8$ grids up
to step $4$. S.M., K.K., and B.W. performed the MPS analysis. All authors
contributed to the writing of the paper.

\section*{Data Availability}

The data supporting the findings of this study are available in
Ref.~\cite{dataset2026}.

\section{Acknowledgments}

We thank Sergio Boixo, Vadim Smelyanskiy, Michael Foss-Feig, and Andrew Potter
for useful discussions and comments.

\bibliography{refs}

\clearpage
\onecolumngrid

\begin{center}
  \textbf{\large Supplement for ``A Heuristic for Matrix Product State Simulation of
  Out-of-Equilibrium Dynamics of Two-Dimensional Quantum Spin Systems''} \\[1.5em]
  Salvatore Mandr\`a,$^{1,*}$ Brayden Ware,$^1$ Nikita Astrakhantsev,$^1$ Sergei Isakov,$^1$ Benjamin Villalonga,$^1$ Tom Westerhout,$^1$ and Kostyantyn Kechedzhi$^1$ \\[0.5em]
  {\small \textit{$^1$Google Quantum AI}} \\[0.5em]
  {\small $^*$smandra@google.com}
\end{center}

\vspace{1em}

\begin{quote}
\small
Out-of-equilibrium dynamics of non-integrable Hamiltonian many-body quantum
systems are characterized by highly entangled wave functions. Near-maximal
entanglement arises in systems exhibiting thermalization or
pre-thermalization, where the system converges to a steady state with a fixed
energy density. Classical simulation of the time dependence of such wave
functions requires exponential resources. However, typical computations aim to
estimate expectation values of local operators and correlation functions to
some expected precision. For thermalizing systems at sufficiently high energy
densities, such computations can be done without storing the full wave
function by instead simulating the evolution of the local operator, which
requires significantly fewer resources.  Nonetheless, constructing such
resource-efficient classical algorithms remains a challenge for intermediate
energy densities, where simulating both the wave function and operator
evolution is costly. In this paper, we propose a heuristic approach to
accelerate the convergence of Matrix Product State (MPS) simulations of
expectation values, applicable across a broad range of energy densities. We
estimate the desired observables by rescaling the MPS results at low bond
dimensions with a factor that depends on the fidelity of the MPS wave
function. Using this technique, we simulated the dynamics of the
two-dimensional Transverse-Field Ising Model (TFIM) on a $7\times8$ grid with
periodic boundary conditions, using a maximum bond dimension of $\chi = 4096$
on a single A100 GPU, as well as the dynamics of the two-dimensional XY model
on grids of size up to $9\times9$. We compare our TFIM results to similar
simulations on a digital quantum processor~\cite{Haghshenas2025Digital},
demonstrating excellent agreement and confirming the predictive power of our
method.
\end{quote}

\vspace{2em}

\setcounter{equation}{0}
\setcounter{figure}{0}
\setcounter{table}{0}
\setcounter{section}{0}

\renewcommand{\theequation}{S\arabic{equation}}
\renewcommand{\thefigure}{S\arabic{figure}}
\renewcommand{\thetable}{S\arabic{table}}
\renewcommand{\thesection}{\Roman{section}}
\def\theHequation{S\arabic{equation}}
\def\theHfigure{S\arabic{figure}}
\def\theHtable{S\arabic{table}}
\def\theHsection{SM.\Roman{section}}

\makeatletter
\renewcommand{\figurename}{Figure}
\renewcommand{\tablename}{Table}
\makeatother

\crefname{figure}{Figure}{Figures}
\Crefname{figure}{Figure}{Figures}
\crefname{table}{Table}{Tables}
\Crefname{table}{Table}{Tables}

\section{Introduction}\label{sec:sm_intro}

This Supplemental Material extends the application of the rescaling method
introduced in the main text to demonstrate its generalizability and robustness
across different physical models and higher dimensions. To probe the breadth
of applicability of this technique for accelerating the convergence of Matrix
Product State (MPS) calculations, here we also consider the dynamics of the
transverse-field Ising model (TFIM) on a three-dimensional lattice. This model
introduces new physical ingredients—such as higher spatial connectivity—thereby
providing a stringent test for the simple rescaling ansatz.

Finally, we summarize the main symbols and notation used throughout both the
main text and this Supplemental Material in \Cref{tab:symbols}.

\section{MPS Ordering}\label{sec:ordering}

In our simulations, we adopt a ``snake-like'' ordering of the MPS sites (or
blocks of sites) for simplicity. However, the efficiency of simulating a 2D
system using an MPS depends on the chosen mapping from the 2D grid to the 1D
MPS chain. This is because any 1D mapping of a 2D grid necessarily introduces
long-range interactions in the MPS representation, which typically accelerates
entanglement growth and requires larger bond dimensions.

A snake-like ordering is a standard choice that minimizes the bandwidth (i.e.,
the maximum distance in the MPS chain between physically adjacent 2D sites)
along one direction (e.g., the width of the grid). This choice makes it
efficient for systems with a small aspect ratio. Alternative orderings, such
as the Hilbert curve, attempt to preserve geometric locality in all
directions, potentially reducing the maximum bandwidth for larger, more
symmetric grids.

While the absolute computational cost and the required bond dimension $\chi$
to reach a given precision depend on the ordering choice, the validity of our
rescaling heuristic is independent of it. The power-law scaling of the
expectation values with respect to the fidelity $F_{\rm MPS}$ stems from the
truncation process and the chaotic nature of the dynamics, rather than the
specific ordering. However, the specific values of the fitted parameters
(such as $\gamma$) depend on the ordering, as it dictates how entanglement is
distributed and truncated.

\section{3D Transverse Field Ising Model}

To further demonstrate the generalizability of the rescaling method presented in
the main text, we apply it to the Transverse-Field Ising Model (TFIM) on a
three-dimensional lattice.  The 3D TFIM presents a significantly more
challenging benchmark for tensor network methods due to the rapid growth of
entanglement with system size and time—a direct consequence of the higher
spatial connectivity. In such regimes, standard MPS simulations are expected to
quickly become inaccurate due
to truncation errors.\\

The system is described by the Hamiltonian in \Cref{eq:hamiltonian} of
the main text, where the interaction sum $\sum_{\langle i,j \rangle}$ extends
over all nearest-neighbor pairs on a 3D cubic grid. In our simulations,
we consider grids
of various sizes with open boundary conditions. This choice demonstrates that
our method also applies to different boundary conditions. To simulate this 3D
system, we employ the same MPS approach as in the main text but utilize a
snake-like ordering with a one-to-one mapping between qubits and MPS sites (see
\Cref{fig:3D_snake}).\\

As in the two-dimensional case, we initialize the system in a product state of
the form given in~\Cref{eq:initial_state} of the main text. To explore
different energy density regimes, we consider three effective temperatures: low
($\Delta\theta = 0$), mid ($\Delta\theta = \pi/9$), and high ($\Delta\theta =
2\pi/9$).\\

Following the procedure described in the main text, we determine the optimal
exponent $\gamma^*$ that minimizes the error in the rescaled expectation value,
as shown in \Cref{tab:3d_gamma_scaling} and \Cref{fig:TFIM_3D_gamma_scaling}.
The linear correlation between $\gamma^*$ and the reduced bond dimension
$\log_2(\chi)/n$ remains a robust feature in 3D, allowing us to extrapolate
the optimal exponent for larger systems. The expected error also follows a
scaling similar to the 2D case, as shown in \Cref{fig:TFIM_3D_rescaling_error}
(the corresponding fitting parameters are reported in
\Cref{tab:3d_scaling_error_params}). The fitting parameters for both
$\gamma^*$ and the expected error are reported in \Cref{tab:3d_gamma_scaling}
and \Cref{tab:3d_scaling_error_params}, respectively.\\

The physical mechanism underlying the success of the rescaling method in 3D is
analogous to the 2D case discussed in the main text. The 3D TFIM is expected to
be even more chaotic due to the increased connectivity. This enhanced chaos
promotes rapid phase randomization among the discarded states. Consequently,
their contributions to the expectation value of local observables tend to cancel
out, justifying the use of the fidelity as a proxy to rescale the retained
component.\\

In \Cref{fig:TFIM_3D_conv_z2_low_mid_high_temp}, we present the convergence of
$\Z{2}$ for a $3\times3\times4$ lattice at the three considered temperatures.
The plots compare both the raw MPS results and the rescaled estimates
$\Z{2}_{\gamma_{\rm fit}}$ against the exact simulations. For the low and mid
temperatures, the rescaled values effectively track the exact results within the
predicted error bounds. In contrast, for the high-temperature regime, we observe
a small but noticeable deviation where the rescaled results fall outside the
estimated maximum error bound, indicating that the method may underestimate the
error at high energy densities. Nonetheless, these results confirm that our
method also works effectively in 3D, successfully overcoming the limitations
that typically cause standard MPS simulations to fail.
Furthermore, in \Cref{fig:TFIM_3D_conv_z2}, we display the results for larger 3D
TFIM systems where exact results are generally unavailable. The figure compares
the unrescaled MPS results (dashed lines) with the rescaled estimates
$\Z{2}_{\gamma_{\rm fit}}$ (orange-circled lines). As the bond dimension
increases, the rescaled results show a clear trend towards convergence, whereas
the unrescaled values exhibit substantial variations. This behavior reinforces
the utility of our method for estimating observables in large-scale 3D
simulations, even when standard MPS calculations fail to converge.

\newpage

\TableSymbols

\TableTFIMGamma

\TableTFIMError

\FigRescalingErrorTwoD

\FigGammaScalingDensity

\FigXYInit

\FigXYGammaScaling

\FigXYConvergence

\FigXYRescalingError

\FigXYErrorValidation

\FigSnake

\FigTFIMThreeDGammaScaling

\FigTFIMThreeDRescalingError

\FigConvergenceLowMidHighTemp

\FigTFIMThreeDConvergence

\end{document}